\pdfoutput=1

\documentclass[11pt]{article}

\usepackage[preprint]{acl}

\usepackage{times}
\usepackage{latexsym}

\usepackage[T1]{fontenc}

\usepackage[utf8]{inputenc}

\usepackage{microtype}

\usepackage{inconsolata}

\usepackage{graphicx}

\usepackage{hyperref}
\usepackage{url}
\usepackage{soul}
\usepackage{pifont}
\usepackage{xcolor}
\usepackage{colortbl}
\usepackage[most]{tcolorbox}
\usepackage{algorithm}
\usepackage{pifont}
\usepackage[noend]{algpseudocode}
\usepackage{amsfonts}
\usepackage{fdsymbol}
\usepackage{graphicx}
\usepackage{subfig}
\usepackage{mathtools}
\usepackage{amsmath}
\usepackage{amsthm}
\usepackage{listings}
\usepackage{wrapfig}
\definecolor{light_gray}{rgb}{.95,.95,.95}
\definecolor{custompurple}{RGB}{93,0,93}
\definecolor{customorange}{RGB}{255,132,6}
\definecolor{customgold}{RGB}{213,177,52}
\definecolor{customblue2}{RGB}{28,205,188}
\definecolor{no_persona_color}{RGB}{152,226,245}
\definecolor{persona_color}{RGB}{193,167,246}
\usepackage{booktabs}
\usepackage{adjustbox}
\usepackage{scalerel}
\usepackage{fontawesome5}
\usepackage{makecell}
\newcommand{\llmlogo}[1]{%
    \adjustbox{valign=m}{\includegraphics[height=1.4ex]{#1}}%
}

\title{Beyond Offline A/B Testing: Context-Aware Agent Simulation for Recommender System Evaluation}

\author{
 \textbf{Nicolas Bougie\textsuperscript{1}},
 \textbf{Gian Marconi Marconi\textsuperscript{1}},
 \textbf{Xiaotong Ye\textsuperscript{1}},
 \textbf{Narimasa Watanabe\textsuperscript{1}}
 \\ \texttt{\{nicolas.bougie,gianmaria.marconi,tony.yip,narimasa.watanabe\}@woven.toyota}\\
\\
 \textsuperscript{1}Woven by Toyota
}

\begin{document}
\maketitle
\begin{abstract}
Recommender systems are central to online services, enabling users to navigate through massive amounts of content across various domains. However, their evaluation remains challenging due to the disconnect between offline metrics and online performance. The emergence of Large Language Model-powered agents offers a promising solution, yet existing studies model users in isolation, neglecting the contextual factors such as time, location, and needs, which fundamentally shape human decision-making. We introduce ContextSim, an LLM agent framework that simulates believable user proxies by anchoring interactions in daily life activities. A life simulation module generates scenarios specifying when, where, and why users engage with recommendations. To align preferences with genuine humans, we model agents’ internal thoughts and enforce consistency at both action and trajectory levels. Experiments across domains demonstrate closer alignment with human behavior than prior work. We further validate our approach through offline A/B testing correlation and show that RS parameters optimized using ContextSim yield improved real-world engagement.
\end{abstract}

\section{Introduction}\label{sec1}
In the era of information explosion, recommender systems (RS) have become an indispensable component of digital platforms, providing personalized recommendations that shape user satisfaction across applications from e-commerce to social media \cite{li2024recent}. Nevertheless, evaluating offline their true effectiveness remains an open challenge \cite{yoon2024evaluating}. Traditional evaluation relies on proxy metrics such as \texttt{hit rate} or \texttt{Recall@N}, but these often fail to predict how users will behave once a system is deployed \cite{zhang2019deep}. The root cause lies in a fundamental disconnect: offline metrics capture static preference patterns, whereas real users make decisions dynamically, influenced by context, internal state, and circumstances \cite{jannach2019measuring}. Online A/B testing addresses this gap but introduces its own drawbacks, including high costs, privacy issues, and ethical concerns around exposing users to potentially suboptimal experiences.

\begin{figure}[tbp]
    \begin{center}
        \includegraphics[width=1.0\linewidth]{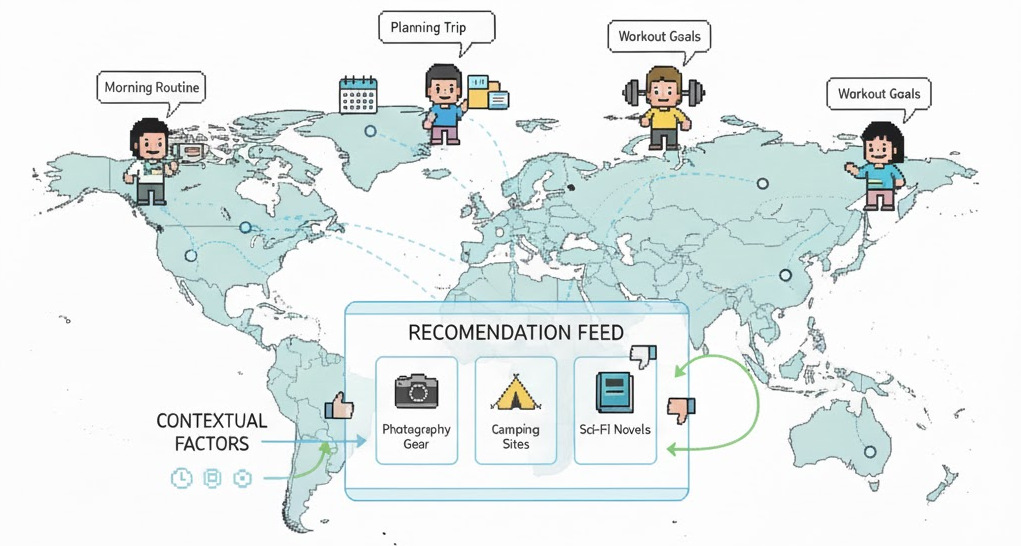}
    \caption{The ContextSim framework for evaluating recommender systems.}
    \label{fig:overall}
\end{center}
\end{figure}

Breakthroughs in Large Language Models (LLMs) have shown promise in human behavior modeling by enabling the creation of autonomous agents. In the realm of recommendation systems, RecMind \cite{wang2023recmind} explores the concept of autonomous recommender agents equipped with self-inspiring planning and external tool utilization. Recently, InteRecAgent \cite{huang2023recommender} has extended this idea by proposing memory components, interactive task planning, and reflection. RecAgent \cite{wang2023user} has attempted to introduce more diverse user behaviors, taking into account external social relationships. Another work, SimUSER \cite{bougie2025simuser}, investigates aligning synthetic agents with their human counterparts through path-driven retrieval and image-grounded reasoning. However, a common drawback of prior studies is their \textit{insulated nature} as they primarily rely on past interactions to make decisions, neglecting environmental factors \cite{Adomavicius2015}. For example, a user shopping in preparation for an upcoming trip may prioritize specific items such as travel accessories, while the same user browsing after a recent move may instead explore furniture or home organization products. Current approaches overlook these contextual factors, leading to synthetic users that behave with unrealistic uniformity. Besides, existing work relies heavily on few-shot exemplars to align agents with historical preferences \cite{lisimulating}. In contrast, we argue that explicitly modeling internal thoughts is essential for generalizing preferences to unfamiliar items and maintaining coherent behavior across time and varying contexts.

In light of this, we introduce ContextSim, a framework that simulates contextualized user behavior and produces faithful interaction trajectories through explicit thought synthesis. Rather than modeling isolated browsing sessions, ContextSim generates daily schedules for each agent, specifying activities, locations, goals, and budgets that shape when, where, and why agents engage with the recommender system. To align users with historical preferences, agents engage in explicit thought modeling: they articulate why a chosen action aligns with their persona, preferences, and situational context. Thought synthesis is performed at both the action and trajectory level via two auxiliary tasks, ensuring internal coherence and belief emergence. Agents then interact with the recommender system following the resulting policy conditioned on their persona and memory modules. This produces fine-grained interaction trajectories that enable reliable RS evaluation and metric estimation under realistic, context-aware settings.

\section{Related Work}
Early work on user simulation relied on bandit-based models that learned from binary feedback signals \cite{christakopoulou2016towards}. KuaiSim \cite{zhao2023kuaisim} advanced this line of research by introducing a simulation platform that supports richer interaction modes. Subsequent approaches incorporated natural language interfaces, but user responses were still largely constrained to predefined choices, such as binary or multiple-choice feedback \cite{lei2020estimation}. As a result, these rule-based simulators lack the behavioral diversity and adaptability observed in real users. The advent of large language models has enabled a new generation of generative user simulators with increased flexibility and expressiveness \cite{zhang2024agentcf}. Such models produce more natural language interactions and reduce reliance on hand-crafted rules, yet challenges remain in calibration, consistency, and behavioral realism. In practice, many LLM-based simulators still depend on scripted interaction flows or weak alignment mechanisms, limiting their ability to exhibit diverse and coherent behaviors over time \cite{lei2020estimation,zhao2023kuaisim}.

\noindent Parallel to this line of work, several studies have explored the use of LLMs as recommender systems themselves, rather than as user simulators \cite{hou2024large,li2023gpt4rec,kang2023llms}. These approaches investigate the capacity of LLMs to generate recommendations directly, providing a perspective complementary to ours, which focuses on modeling how users perceive and respond to recommendations \cite{wang2024llm,zhang2024prospect}.

\noindent More closely related to our work are LLM-based autonomous user agents for recommendation simulation. RecMind \cite{wang2023recmind} pioneers self-inspiring agents but limits interactions to simple rating actions. Yoon et al.~\cite{yoon2024evaluating} conduct a systematic study of LLM effectiveness in conversational recommendation simulation. Agent4Rec \cite{zhang2023generative} introduces memory mechanisms to maintain interaction history and improve behavioral faithfulness. Recently, RecInter \cite{jin2025beyond} seeks to reproduce phenomena such as Matthew Effect. SimUSER \cite{bougie2025simuser} further advances this direction by incorporating persona matching via self-consistency scoring, knowledge-graph memory to capture user--item relationships, and visual perception for thumbnail-driven decisions. Similarly, AlignUSER leverages counterfactual reasoning and next-state prediction to make agents aware of the world \cite{bougie2026alignuser}. Despite these advances, existing approaches predominantly rely on weak alignment signals, such as few-shot prompting, and model user behavior largely in isolation from the broader situational context. Meanwhile, the importance of context for recommendation quality has long been established \cite{adomavicius2011context}: temporal dynamics influence preference expression \cite{koren2009collaborative}, location affects perceived relevance \cite{liu2017experimental}, and situational factors shape decision criteria \cite{zheng2015carskit}. However, this rich literature on context-aware recommendation has not yet translated into context-aware user simulation.

\section{Method}
ContextSim presents a framework for evaluating RS through contextually-grounded agent interactions. Our approach comprises three phases: (1) persona initialization from historical data, (2) thought modeling via explicit reasoning, (3) daily life simulation that generates environmental context shaping subsequent interactions with the RS. 

\noindent \textbf{Problem Formulation:} We model user–recommender interaction as a tuple $\mathcal{M} = (\mathcal{S}, \mathcal{A}, T)$, where $\mathcal{S}$ denotes the state space and $\mathcal{A}$ is the action space. The transition function $T\colon \mathcal{S} \times \mathcal{A} \rightarrow \mathcal{S}$ governs the environment dynamics. In our study, the state $s$ encodes what is visible to the agent (e.g., a ranked list of items $i \in \mathcal{I}$) and context $c \in \mathcal{C}$ (e.g., temporal, spatial, and situational factors). An action $a_t \in \mathcal{A}$ represents discrete operations such as inspecting an item, clicking, adding to cart, rating, or exiting. Our objective is to simulate realistic trajectories $(s_0, a_0, \dots, s_T)$ for each user following a policy $\pi_{\theta} : S \rightarrow A$ parameterized by $\theta$, including predicting the rating $r_{ui}$ for unseen items $i$ given $c$. 

\subsection{Persona Initialization}
We represent each agent's profile $p$ through demographic and psychological attributes: \textbf{age}, \textbf{occupation}, and \textbf{personality}. Personality follows the Big Five traits: $\in$ \{\textit{Openness}, \textit{Conscientiousness}, \textit{Extraversion}, \textit{Agreeableness}, \textit{Neuroticism})\}, scored on a 1-3 scale. Since real-world datasets rarely include such attributes, we infer them from interaction histories using self-consistent persona selection \cite{bougie2025simuser}. We first prompt the LLM to generate plausible candidate personas, then score via a consistency score. The persona with the highest consistency is assigned to the agent. Beyond these static attributes, $p$ also includes: \textbf{habits}, \textbf{recent goals}, \textbf{preferences}. Habits account for user tendencies in engagement, conformity, and variety \cite{zhang2023generative}. Recent goals are inferred from prior sessions, and preferences are short natural-language descriptions that summarize the user's tastes. 

\noindent\textbf{Memory Module} 
The episodic memory records the interactions with the RS. This memory is initially populated with the user's viewing, rating history, and preferences. Each time an agent executes a new action or rates an item, the corresponding interaction is added to the episodic memory. We also maintain an emotional memory that records user feelings during system interactions, such as level of fatigue and satisfaction, capturing the psychological feeling stemming from past interactions.

\subsection{Thought Synthesis}
Relying solely on persona-style instructions (e.g., \textit{``act as \dots''}) often leads LLM agents to overfit persona descriptors rather than understanding their actual interaction patterns. To better align agents with user preferences and enable generalization beyond previously seen items, we argue that explicitly modeling the user's internal thoughts is crucial. To do so, we let agents synthesize their own thoughts, such as \textit{I loved The Matrix and Blade Runner. Interstellar has similar sci-fi themes and Nolan directed it, so I'm clicking to watch.}. Given a user's interaction history $H \leftarrow \{(s_{t}, a_{t}, c, p, H_{a}), ...\}$, the agent compares the candidate actions and infers why the human choice aligns with $c$, $p$, and past actions $H_{a}$. Concretely, we introduce two reasoning tasks at the item and trajectory levels.

\noindent\textbf{Item Disentanglement.} Given an item $i$ and the user's historical action $a_{t}$ (e.g., purchase, click, rating), the agent infers why the action reflects its preferences. The agent is instructed to ground the explanation in persona $p$, interaction history $H$, and item features.

\noindent\textbf{Trajectory Alignment:} In this task, the agent is given a state $s_{t}$, the historical action $a_{t}$, alternative navigation actions $A_{t}$ (e.g., next page, exit), history $H$, and is prompted to generate a brief rationale that: (i) why the historical action is preferred over other actions, and (ii) highlights why the action aligns with $p$ and $H$. The prompt explicitly requires the thought to reference observable aspects rather than generic justifications. 

\noindent The resulting chain-of-thought $c_{ID}$ and $c_{TA}$ are collected into a dataset $\mathcal{D}_{ID}$ and $\mathcal{D}_{TA}$ respectively. We train $\pi_{\theta}$ via SFT with a joint objective over reasoning generation:
\begin{equation}
\resizebox{\columnwidth}{!}{$
\begin{aligned}
\mathcal{L}_{\text{ST}} &=
- \sum_{(s_t, a_{t}, c, p, H_{a}, c_{ID}) \in \mathcal{D}_{\text{ID}}}
  \log p_{\theta}\!\bigl(c_{ID} \mid s_t, a_{t}, c, p, H_{a}\bigr) \\
&\quad
- \sum_{(s_t, a_{t}, A_{t}, c, p, H_{a}, c_{TA}) \in \mathcal{D}_{\text{TA}}}
  \log p_{\theta}\!\bigl(c_{TA} \mid s_t, a_{t}, A_{t}, c, p, H_{a}\bigr)
\end{aligned}
$}
\end{equation}

\subsection{Life Simulation Module}
\label{sec:context}
In the real world, interactions do not occur in isolation but emerge from the activities and constraints of everyday life. Thus, we utilize a life simulation module \cite{bougie2025citysim} to produce realistic contexts. For each agent with persona $p$, we generate a daily schedule $\mathcal{S} = \{(t_1, a_1, l_1), \ldots, (t_n, a_n, l_n)\}$ where each tuple represents a time slot $t$, activity $a$, and location $l$. The schedule is conditioned on persona attributes, day type, past activities, and external factors (weather conditions, local events, season).

\noindent From these schedules, we identify when recommendation interactions would naturally occur and construct a corresponding contextual scenario $c$ comprising:
\begin{itemize}\setlength\itemsep{0pt}\setlength\parskip{0pt}
    \item \textbf{Temporal context} $c_t$: time of day, day of week,
    \item \textbf{Spatial context} $c_l$: current location,
    \item \textbf{Situational context} $c_s$: latest activity, mood, need level (e.g., hunger), and energy level,
    \item \textbf{Goal context} $c_g$: purpose of seeking recommendations,
    \item \textbf{Constraint context} $c_b$: budget, time availability,
\end{itemize}
The full context vector is: $c = (c_t, c_l, c_s, c_g, c_b)$.

\subsection{Context-Aware Interactions}
We now consider two ways of interacting with the RS: \textsc{ContextSim(sim)}, \textsc{ContextSim(sum)}. In the former, at each step, we run the life simulation module, simulate the context, and if the agent decides to seek recommendations, it engages with the RS. In the latter, we run the life simulation module for 30 days, then summarize contextual factors and append them to the agent's persona. 

\noindent When interacting with the RS, the agent conditions its decisions on the current context~$c$. It first senses the page and evaluates each item by estimating its alignment with the persona, context, and retrieved evidence from the episodic memory, producing a shortlist of \texttt{[WATCH]} or \texttt{[SKIP]} intentions. Following this, the agent infers its fatigue, curiosity, and boredom, before action selection. When selecting an action, the agent generates an internal thought that weighs the available actions (e.g., scroll, click, add to cart, rate, exit) against its goals and constraints. Each reasoning step is accompanied by a \textit{thought} to maintain behavioral consistency. This loop is repeated until it decides to stop interacting with the RS for the current session or selects a final action (e.g., \texttt{[PURCHASE CART]}, \texttt{[EXIT]}). After acting, the agent performs self-reflection, updating the memory module with concise reflections explaining its decisions and evolving tastes.

\section{Experiments}

\noindent\textbf{Settings.} We evaluate \textsc{ContextSim} with Qwen3-8B \cite{yang2025qwen3} as backbone. Unless otherwise specified, we report results obtained with the summary-based variant, \textsc{ContextSim(sum)}. All experiments are conducted with 1,000 agents.

\noindent\textbf{Datasets.} Evaluation spans four domains: MovieLens-1M (movies), AmazonBook (books), and Steam (video games), and OPeRA \cite{wang2025opera} (EC site). We follow standard preprocessing and use time-based 80/10/10 splits for train/validation/test. For temporal analysis, we leverage MovieLens and OPeRA timestamps to establish ground-truth interaction patterns. Besides, for trajectory-level evaluation, we rely on the OPeRA dataset \cite{wang2025opera}.

\noindent\textbf{Baselines.} We compare our method against RecAgent, Agent4Rec, and SimUSER as LLM-powered agent baselines. When possible, we also include RecMind results. All baselines are implemented using GPT-4o-mini, as reported in their original experimental settings.

\subsection{Preference Alignment}
\label{sec:beliaviability}
For synthetic users to provide useful feedback to a recommender system, they should preserve the preference patterns of the real users they represent. We therefore prompt the agents to classify items based on whether their human counterparts have interacted with them or not. For each of 1{,}000 agents, we sample 20 candidate items and vary the ratio of interacted to non-interacted items as 1:$m$, where $m \in \{1,3,9\}$ and non-interacted items are labeled as $y_{ui}=0$. This yields a binary classification task over user--item pairs. Table \ref{table:taste_alignment} highlights that ContextSim agents accurately identified items aligned with their tastes, significantly outperforming baselines across all distractor levels (paired t-tests, 95\% confidence, $p < 0.001$). The improvement stems from modeling thoughts, as agents can understand what features or attributes (e.g., genre, brand, price range) drive their preferences, enhancing alignment.

\begin{table*}[tb]
\centering
\begin{adjustbox}{width=\textwidth}
\begin{tabular}{cccccccccccccc}
\toprule
 & \multicolumn{4}{c}{MovieLens} & \multicolumn{4}{c}{AmazonBook} & \multicolumn{4}{c}{Steam} \\ 
\cmidrule(lr){2-5} \cmidrule(lr){6-9} \cmidrule(lr){10-13}
Method(1:m) & Accuracy & Precision & Recall & F1 Score & Accuracy & Precision & Recall & F1 Score & Accuracy & Precision & Recall & F1 Score \\ 
\midrule
RecAgent (1:1) & 0.5807 & 0.6391 & 0.6035 & 0.6205 & 0.6035 & 0.6539 & 0.6636 & 0.6587 & 0.6267 & 0.6514 & 0.6490 & 0.6499 \\
RecAgent (1:3) & 0.5077 & 0.7396 & 0.3987 & 0.5181 & 0.6144 & 0.6676 & 0.4001 & 0.5003 & 0.5873 & 0.6674 & 0.3488 & 0.4576 \\
RecAgent (1:9) & 0.4800 & 0.7491 & 0.2168 & 0.3362 & 0.6222 & 0.6641 & 0.1652 &  0.2647 & 0.5995 & 0.6732 & 0.1744 & 0.2772 \\
\midrule
Agent4Rec (1:1) & 0.6912 & 0.7460 & 0.6914 & 0.6982 & 0.7190 & 0.7276 & 0.7335 & 0.7002 & 0.6892 & 0.7059 & 0.7031 & 0.6786 \\
Agent4Rec (1:3) & 0.6675 & 0.7623 & 0.4210 & 0.5433 & 0.6707 & 0.6909 & 0.4423 & 0.5098 & 0.6505 & 0.7381 & 0.4446 & 0.5194 \\
Agent4Rec (1:9) & 0.6175 & 0.7753 & 0.2139 & 0.3232 & 0.6617 & 0.6939 & 0.2369 & 0.3183 & 0.6021 & 0.7213 & 0.1901 & 0.2822 \\
\midrule
SimUSER (1:1) & 0.7912 & 0.7976 & 0.7576 & 0.7771 & 0.8221 & 0.7969 & 0.7841 & 0.7904 & 0.7905 & 0.8033 & 0.7848 & 0.7939 \\
SimUSER (1:3) & 0.7737 & 0.8173 & 0.5223 & 0.6373 & 0.6629 & 0.7547 & 0.5657 & 0.6467 & 0.7425 & 0.8048 & 0.5376 & 0.6446 \\
SimUSER (1:9) & 0.6791 & 0.8382 & 0.3534 & 0.4972 & 0.6497 & 0.7588 & 0.3229 & 0.4530 & 0.7119 & 0.7823 & 0.2675 & 0.3987 \\
\midrule
ContextSim (1:1) & 0.824 & 0.831 & 0.789 & 0.809 & 0.851 & 0.823 & 0.812 & 0.817 & 0.819 & 0.828 & 0.801 & 0.814 \\
ContextSim (1:3) & 0.798 & 0.841 & 0.551 & 0.666 & 0.689 & 0.779 & 0.592 & 0.673 & 0.768 & 0.829 & 0.561 & 0.669 \\
ContextSim (1:9) & 0.712 & 0.861 & 0.381 & 0.528 & 0.678 & 0.784 & 0.351 & 0.485 & 0.739 & 0.809 & 0.294 & 0.431 \\
\bottomrule
\end{tabular}
\end{adjustbox}
\caption{User preference alignment across MovieLens, AmazonBook, and Steam datasets.}
\label{table:taste_alignment}
\end{table*}

\subsection{Rating Items}

\begin{table}[tbp]
\centering
\resizebox{1.0\linewidth}{!}{
\begin{tabular}{lcccccc}
\toprule
\textbf{Methods} & \multicolumn{2}{c}{\textbf{MovieLens}} & \multicolumn{2}{c}{\textbf{AmazonBook}} & \multicolumn{2}{c}{\textbf{Steam}} \\
 & \textbf{RMSE} & \textbf{MAE} & \textbf{RMSE} & \textbf{MAE} & \textbf{RMSE} & \textbf{MAE} \\
\midrule
MF & 1.214 & 0.997 & 1.293 & 0.988 & 1.315 & 1.007 \\
AFM & 1.176 & 0.872 & 1.301 & 1.102 & 1.276 & 0.972 \\
RecAgent & 1.102 & 0.763 & 1.259 & 1.119 & 1.077 & 0.960 \\
Agent4Rec & 0.761 & 0.714 & 0.879 & 0.671 & 0.758 & 0.688 \\
SimUSER & 0.502 & 0.446 & 0.568 & 0.421 & 0.587 & 0.532 \\
\midrule
\rowcolor{blue!10}
ContextSim & \textbf{0.451} & \textbf{0.392} & \textbf{0.511} & \textbf{0.369} & \textbf{0.528} & \textbf{0.471} \\
\phantom{  } ContextSim(fs) & \underline{0.488} & \underline{0.421} & \underline{0.542} & \underline{0.408} & \underline{0.561} & \underline{0.503} \\
\bottomrule
\end{tabular}
}
\caption{Rating prediction performance. \textbf{Bold}: best results; \underline{underlined}: second-best. ContextSim's improvements are statistically significant ($p< 0.05$).}
\label{fig:rating_prediction}
\end{table}
A fundamental task when interacting with a RS is rating items. We compare several LLM-based baselines, along with traditional recommendation baselines: MF \cite{koren2009matrix} and AFM \cite{xiao2017attentional}. We also compare against a GPT-4o-mini variant, denoted \textsc{ContextSim(fs)}, which replaces thought-synthesis pretraining with thoughts as few-shot exemplars. Across all tasks (Table \ref{fig:rating_prediction}), ContextSim outperforms other LLM-powered agents. Notably, the thought-synthesis–trained model, \textsc{ContextSim}, surpasses the few-shot variant \textsc{ContextSim(fs)}, even when the latter relies on a substantially larger LLM. This gap highlights an inherent limitation of few-shot prompting: without explicit training on structured reasoning, the model fails to internalize how preferences, context, and history jointly shape user ratings, resulting in shallow and often brittle decision behavior.

\subsection{Thought Consistency}
\begin{table}[tbp]
\resizebox{1.0\linewidth}{!}{
\centering
\begin{tabular}{lcc}
\toprule
\textbf{Method} & \textbf{Consistency Rate} & \textbf{Contradiction Rate} \\
\midrule
RecAgent & 17.3\% & 38.2\% \\
Agent4Rec & 21.8\% & 34.6\% \\
SimUSER & 29.2\% & 29.8\% \\
\rowcolor{blue!10}
ContextSim & \textbf{84.1\%} & \textbf{5.3\%} \\
\bottomrule
\end{tabular}}
\caption{Persona-action consistency rates evaluated by GPT-4o. Improvements are statistically significant (p < 0.05).}
\label{tab:consistency}
\end{table}
We next assess whether explicit thought modeling helps agents maintain persona-consistent behavior. We use the OPeRA dataset~\cite{wang2025opera}, which provides shopping sessions annotated with user personas, step-level rationales, and actions. For each rationale, the agent receives the same persona, the observation, and the interaction history, and is prompted to generate both an internal rationale and a next action. Following prior work, GPT-4o acts as an automatic judge and labels each (rationale, action) pair as \emph{coherent}, \emph{partially coherent}, or \emph{contradictory} with respect to the persona and context. We report the proportion of coherent steps as the consistency rate and the proportion of contradictory steps as the contradiction rate. As shown in Table~\ref{tab:consistency}, ContextSim demonstrates a substantially higher consistency rate than all baselines, while reducing contradiction.

\subsection{Believability of Synthetic Users}
\label{sec:believability}
\begin{table}[btp]
\centering
\resizebox{1.0\columnwidth}{!}{
\begin{tabular}{lcccc}
\toprule
 & \textbf{MovieLens} & \textbf{AmazonBook} & \textbf{Steam} & \textbf{OPeRA} \\
\midrule
RecAgent & 3.01 $\pm$ 0.14 & 3.14 $\pm$ 0.13 & 2.96 $\pm$ 0.17 & 3.05 $\pm$ 0.15 \\
Agent4Rec & 3.04 $\pm$ 0.12 & 3.21 $\pm$ 0.14 & 3.09 $\pm$ 0.16 & 3.15 $\pm$ 0.17 \\
SimUSER & 4.41$\pm$0.16 & 3.99$\pm$0.18 & 4.02$\pm$0.23 & 4.05$\pm$0.20 \\
\rowcolor{blue!10}
ContextSim(sim) & 4.58 $\pm$ 0.13* & \textbf{4.32 $\pm$ 0.17}* & 4.26 $\pm$ 0.21* & 4.21$\pm$0.20*\\
\rowcolor{blue!10}
ContextSim(sum) & \textbf{4.60 $\pm$ 0.14}* & 4.28 $\pm$ 0.16* & \textbf{4.31 $\pm$ 0.19}* & \textbf{4.21$\pm$0.18}*\\
\bottomrule
\end{tabular}}
\caption{Human-likeness score evaluated by GPT-4o across recommendation domains. *Significant improvements over best baseline ($p<0.05$).}
\label{tab:llm_evaluator}
\end{table}

Following prior work on LLM-based evaluators~\cite{chiang2023can, bougie2025citysim}, which has shown that LLM judges can approximate human preferences, we use GPT-4o to assess whether agent interactions appear human or AI-generated using a 5-point Likert scale, with higher scores indicating stronger resemblance to human-like responses. As reported in Table~\ref{tab:llm_evaluator}, ContextSim achieves the highest scores across all four domains, with statistically significant improvements over SimUSER in every dataset ($p<0.05$). We noticed that the life-context grounding contributes to the faithfulness of our method, especially in the generated rationales. In addition, explicit thought synthesis reduces overly generic or uniformly positive reactions that exhibit prior work.

\subsection{Offline A/B Testing}
\begin{figure}[tbp]
    \centering
    \includegraphics[width=1.0\linewidth]{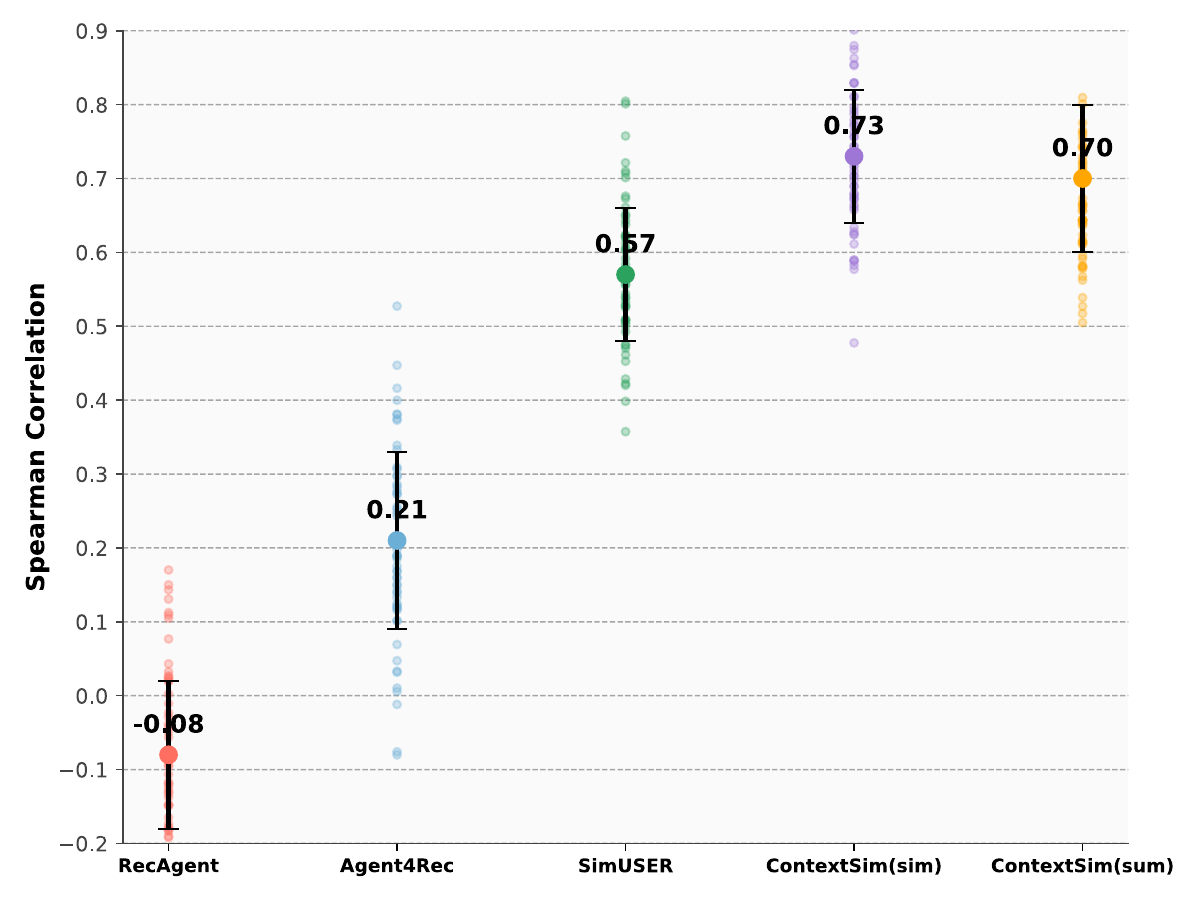} 
    \caption{Spearman correlation between estimated and actual engagement metrics. Higher values indicate better alignment with ground-truth metrics.}
    \label{fig:ab_tests}
\end{figure}

We further examine whether ContextSim can serve as a reliable proxy for online A/B tests. We use a proprietary dataset of 55 historical A/B experiments on a large-scale food recommendation platform, each involving thousands of user sessions and hundreds of thousands of recommended item impressions. Every test compares multiple recommendation strategies, with the average number of visited pages used as the primary business metric. For each historical test, we instantiate one agent for each user in the corresponding A/B population and initialize the agent only with information available before the test period, such as past interactions and preference history. For each strategy, we run the corresponding simulator and estimate the same engagement metric, then compute the Spearman correlation between simulated and real-world outcomes across the 55 tests. As illustrated in Figure~\ref{fig:ab_tests}, our approach achieves the highest correlation with ground truth, outperforming all other baselines. Statistical tests confirm that the improvements over all baselines are significant ($p<0.05$).

\subsection{Optimizing RS Parameters}

\begin{table}[tbp]
\centering
\resizebox{1.0\columnwidth}{!}{
\begin{tabular}{lccccc}
\toprule
\textbf{Method} & $\overline{P}_{\text{view}}$ & $\overline{N}_{\text{like}}$ & $\overline{P}_{\text{like}}$ & $\overline{N}_{\text{exit}}$ & $\overline{S}_{\text{sat}}$ \\
\midrule
Baseline & 0.521 & 5.44 & 0.458 & 3.21 & 3.82 \\
Traditional (nDCG@10) & 0.535 & 5.52 & 0.462 & 3.26 & 3.86 \\
SimUSER-optimized & 0.561 & 5.80 & 0.517 & 3.87 & 4.09 \\
ContextSim-optimized & \textbf{0.589} & \textbf{6.12} & \textbf{0.543} & \textbf{4.15} & \textbf{4.24} \\
\bottomrule
\end{tabular}}
\caption{Performance comparison of parameter selection strategies on various engagement metrics. }
\label{tab:param_selection}
\end{table}
We now assess whether choosing RS parameters based on ContextSim(sim)’s evaluations leads to measurable gains in a real deployment. ContextSim is used to rank candidate parameter settings before deployment, and Table~\ref{tab:param_selection} reports the metrics observed after deploying the selected settings in the real world. Namely, the table compares the baseline RS, with three selection strategies: traditional offline metrics (nDCG@10), SimUSER, and ContextSim. Results depicted in the table are computed from real logs collected after deploying the \textit{optimized} recommender system. Optimizing for nDCG@10 yields only marginal improvements over the baseline, echoing prior findings that offline accuracy metrics do not reliably translate into business impact~\cite{jannach2019measuring}. In contrast, parameters selected using ContextSim achieve the best results across all engagement metrics, including higher average viewing ratio ($\overline{P}_{\text{view}}$), more liked items ($\overline{N}_{\text{like}}$), and higher satisfaction scores ($\overline{S}_{\text{sat}}$). Compared to SimUSER-optimized parameters, ContextSim yields a higher viewing ratio and satisfaction, indicating that richer contextual grounding provides more actionable guidance for real-world RS tuning.

\section{Conclusion}
We propose \textsc{ContextSim}, a simulation framework that grounds user agents in realistic daily life contexts. Rather than relying on few-shot prompting, we pretrain agents through thought synthesis tasks that teach them to articulate why their decisions align with their persona, history, and situational context. Across multiple datasets and evaluation metrics, \textsc{ContextSim} produces more believable, context-aware trajectories than existing techniques. Our agents align closely with human preferences, capture temporal interaction patterns overlooked by prior work, and exhibit a strong correlation with real-world A/B test outcomes. Beyond mere correlation, we demonstrate that RS parameters optimized using our approach translate directly into higher real-world KPIs, confirming that user simulation can drive measurable improvements in production systems.

\section{Ethics Statement}
This work proposes an LLM-driven agent framework that simulates user interactions with recommender systems. While synthetic users can reduce reliance on costly or intrusive human experiments, they also raise ethical concerns. Detailed daily life simulations could, if misused, enable fine-grained profiling, behavioral manipulation, or the reinforcement of harmful stereotypes.

\noindent We emphasize that ContextSim is intended for \emph{system-level} evaluation and analysis, not for targeting or inferring attributes of specific individuals. All datasets used in our experiments are either public benchmarks or proprietary logs that were anonymized and aggregated before analysis, following the data governance policies of our organization. The explicit thought modeling in ContextSim provides a degree of transparency: it makes it easier to inspect and understand long-tail behaviors, rather than hiding them inside a black-box policy.

\noindent We view synthetic users as a complement, not a replacement, for real user feedback, especially for populations that may be underrepresented or poorly modeled by LLMs. Practitioners deploying frameworks like ContextSim should ensure that their use complies with applicable privacy regulations, includes regular audits of potential biases, and incorporates human oversight in high-stakes decisions.

\section{Limitations}
Although ContextSim achieves strong performance across the evaluated metrics, several limitations remain. First, reproducibility is constrained by the fact that part of our evaluation relies on one proprietary interaction dataset. Second, as with all LLM-based simulators, ContextSim may inherit cultural, gender, and socioeconomic biases present in foundation models. We also observe occasional hallucinations, particularly when the model generates appraisals for rare or recently added or popular items, which can introduce noise into simulated interactions. Moreover, the fidelity of the simulated behavior ultimately depends on the capabilities and failure modes of the underlying LLMs. Inconsistent reasoning, biased interpretations, or unfounded assumptions made by the base model can propagate through the simulation pipeline. Another limitation is that part of the framework remains prompt-dependent. Although the interaction policy is fine-tuned with decision-rationale supervision, modules such as persona initialization, life simulation, and reflection may remain sensitive to prompt design. We evaluate prompt robustness in Section~\ref{sec:prompt_robustness}. Nevertheless, transferring the framework to substantially different domains or interfaces may require revising these prompts and validating their effects. Finally, ContextSim integrates multiple interacting modules (life simulation, thought modeling), making it challenging to attribute improvements to a single component. While our ablation studies attempt to evaluate the contribution of each module, fully isolating their effects remains difficult.

\bibliography{custom}
\clearpage
\appendix
\section{Experimental Setup}
Experiments are conducted on four datasets spanning diverse recommendation domains: MovieLens-1M (movies), AmazonBook (books), Steam (video games), and OPeRA~\cite{wang2025opera} (EC site). Following prior work, we filter out users and items with fewer than 20 interactions. We then split interactions into training, validation, and test sets using an 80/10/10 time-based split, so that all test interactions strictly occur after the validation and training interactions. This reflects the temporal distribution shift encountered in real deployments and ensures that simulated agents cannot access future information. In a few experiments, we also report results for \textsc{ContextSim(fs)}, where agents are not pretrained but instead condition their decisions on chain-of-thoughts exemplars appended to prompts as structured \textit{context}. This enables a direct comparison between pretrained agents and those that rely purely on prompt interpretation and in-context reasoning.

\noindent\textbf{Agent Initialization.} For each user, we instantiate one ContextSim agent. The agent profile $p$ is initialized via the persona-matching procedure described in Section~\ref{sec:context}. Concretely, we provide the LLM a few interactions from its history and prompt it to generate $K=5$ candidate personas. A separate scoring prompt evaluates the consistency between each candidate persona and the interaction history; the persona with the highest consistency score is selected and fixed for all subsequent simulations. Episodic memory is initialized with the user's training interactions and preference descriptions, while values in the emotional memory are initialized as neutral and updated during simulations.

\noindent\textbf{Thought Synthesis.} We construct two reasoning datasets, $\mathcal{D}_{\text{PD}}$ (item disentanglement) and $\mathcal{D}_{\text{TA}}$ (trajectory alignment), from the training splits of all datasets. For $\mathcal{D}_{\text{PD}}$, we sample rated/liked/purchased items for each user (up to 50 per user) and prompt the model to explain why the observed action is consistent with the persona $p$, history $H$, and item attributes. For $\mathcal{D}_{\text{TA}}$, we use state-action pairs and alternative navigation actions $A_t$, and ask the model to justify why the historical action $a_t$ is preferable, grounding the explanation in $p$, $H$, and the current state $s_t$. We fine-tune a Qwen3-8B model on $\mathcal{D}_{\text{PD}} \cup \mathcal{D}_{\text{TA}}$. Unless otherwise stated, we train for 5 epochs with a batch size of 16, learning rate $1\!\times\!10^{-5}$, AdamW optimizer, and a maximum sequence length of 4{,}096 tokens. LoRA was applied to all linear modules, while LoRA rank is set to 8 and LoRA alpha is set to 16. 

\noindent\textbf{Life Simulation.} The life simulation module is instantiated using the personas inferred from the training split and is applied independently to each agent. Prompts follow CitySim framework \cite{bougie2025citysim}. For \textsc{ContextSim(sim)}, we simulate day-by-day schedules during evaluation, discretizing the day into 30-minute slots and conditioning the generated activities on persona attributes, day type (weekday vs. weekend), and sampled external factors (season, weather, local events). At each slot, the module determines whether an RS interaction is likely to occur; if so, it produces a contextual scenario $c = (c_t, c_l, c_s, c_g, c_b)$, which is passed to the interaction policy. For \textsc{ContextSim(sum)}, we run the life simulation for 30 days per agent before evaluation and summarize contextual statistics (e.g., typical time-of-day and location of usage, recurring goals and constraints) into a short text description that is appended to the persona. All subsequent interactions for \textsc{ContextSim(sum)} are conditioned on this summary instead of running the life simulation online.

\noindent\textbf{Tasks and Metrics.} For rating prediction tasks, we ask agents to assign ratings to held-out user–item pairs and compute RMSE and MAE on the test split. For temporal-pattern analysis, we group interactions into four time-of-day bands (Morning, Afternoon, Evening, Night) using the original timestamps and report click-through rates and Spearman correlations against real data. For action-alignment experiments on OPeRA, we follow the protocol described in Section~\ref{tab:action_predict}, requiring exact match of action parameters and reporting accuracy and F1 metrics over action types, click subtypes, and session outcomes.

\noindent\textbf{Baselines.} We compare ContextSim to RecAgent, Agent4Rec, and SimUSER as LLM-powered user simulators, and to RecMind when results are available. All LLM-based baselines are implemented with GPT-4o-mini as in their original papers and use the same training/validation/test splits as ContextSim. For rating prediction, we also include standard RS baselines: Matrix Factorization (MF) and Attentional Factorization Machines (AFM). GPT-4o serves as an automatic judge for persona–action consistency, human-likeness scores, and context-consistency labels. For the offline A/B testing experiments, we use a proprietary dataset of 55 historical experiments from a large-scale food recommendation platform; for each strategy, we run the simulator on the same item pools and compute the Spearman correlation between simulated and real metrics.

\noindent\textbf{Interactions with RS.} Following persona initialization and thought synthesis, agents interact with the recommender system in a page-by-page manner until a terminal action is selected. In recommendation domains (MovieLens, Steam, AmazonBook), sessions terminate via \texttt{[EXIT]}; while in the web-shopping domain (OPeRA), it may include purchase-related decisions before \texttt{[EXIT]}.  To improve accuracy, they first assess items on the current pages by assigning to each a \texttt{[WATCH]}/\texttt{[SKIP]} intention. Next, the agent infers internal state, including fatigue, curiosity, and boredom, from the ongoing session, recent actions, and current context. It then selects an action from the environment action space (e.g., navigating pages, clicking an item, searching, rating, or exiting). Action selection is accompanied by an internal thought that explicitly weighs the available actions against its persona and context $c$, as well as the agent's preferences and prior interactions. After executing the action, the simulator transitions to the next state $s_{t+1}$, and the new interaction is appended to episodic memory. Following each step, the agent performs a short self-reflection that summarizes the rationale behind the action (and any expressed tastes), storing this concise explanation in the episodic memory.

\section{Discussion}
Our study shows that grounding LLM-based user agents in a realistic daily-life context, along with explicit thought synthesis, yields synthetic users that more closely match human preferences and interaction dynamics, and provides more reliable evaluation signals for recommender systems. Despite these improvements, several limitations remain.

\noindent First, the current framework represents context and interface states primarily through text. While this abstraction enables a simplified implementation across domains, it may omit fine-grained cues that drive real decisions, such as item thumbnails or user experience. Extending ContextSim to multimodal observations and structured interface representations (e.g., interface screenshot) could improve both the realism of user actions.

\noindent Second, our thought-synthesis supervision is derived from historical interactions and their associated outcomes, which provide only partial coverage of the space of plausible contexts and behaviors. Although the life simulation module increases diversity by exposing agents to a broader range of temporal, spatial, and situational conditions, the induced contexts are still generated by a model and may fail to capture rare events, abrupt preference shifts, or hard-to-capture constraints present in real life (e.g., unexpected schedule changes, social commitments). Improving the realism of simulated contexts remains an important research direction.

\noindent Third, our life simulation module currently summarizes context over a fixed horizon and uses either online generation (\textsc{ContextSim(sim)}) or an offline summary (\textsc{ContextSim(sum)}). While the summary variant is computationally efficient and performs well empirically, it can blur short-term fluctuations that affect decisions at the session level (e.g., late-night fatigue or time-limited shopping). Conversely, the online variant increases fidelity but introduces additional variance and may require more interactions to faithfully reproduce human action distributions.

\begin{table*}[tbp]
\centering
\small
\resizebox{0.98\linewidth}{!}{
\begin{tabular}{llllc}
\toprule
\textbf{Evaluation} & \textbf{Section} & \textbf{Metric} & \textbf{Basis} & \textbf{LLM judge} \\
\midrule
Preference Alignment 
& 4.1 
& Acc. / Prec. / Rec. / F1 
& Real interactions (binary labels) 
& No \\

Rating Prediction 
& 4.2 
& RMSE / MAE 
& Held-out ratings 
& No \\

Thought Consistency 
& 4.3 
& Consistency / Contradiction 
& GPT-4o judgment 
& Yes \\

Human-likeness 
& 4.4 
& Likert score 
& GPT-4o judgment 
& Yes \\

Offline A/B Testing 
& 4.5 
& Spearman correlation 
& Real A/B test outcomes (55 experiments) 
& No \\

RS Parameter Optimization 
& 4.6 
& Engagement metrics 
& Real deployment results 
& No \\

Action Alignment (OPeRA) 
& App.~F.5 
& Accuracy / F1 
& Ground-truth actions (exact match) 
& No \\

Context Consistency 
& App.~F.6 
& Alignment labels 
& GPT-4o judgment 
& Yes \\

Temporal Patterns 
& App.~F.8 
& CTR / Correlation 
& Real timestamps 
& No \\

Human Preferences 
& App.~F.12 
& Preferences 
& Human judgment 
& No \\
Rating Distribution 
& App.~F.7 
& Distribution alignment 
& Ground-truth rating distribution 
& No \\

Prompt Robustness 
& App.~F.15 
& RMSE / Action F1 / Temporal corr. 
& Held-out ratings, ground-truth actions, and real timestamps 
& No \\

\bottomrule
\end{tabular}
}
\caption{Summary of the evidence used in each evaluation. }
\label{tab:evaluation_grounding}
\end{table*}

\noindent Fourth, we evaluate \textsc{ContextSim} using several sources of evidence. On MovieLens-1M, AmazonBook, and Steam, we test whether agents recover users' past preferences and predict ratings. On OPeRA, we evaluate whether agents reproduce human web-shopping actions and session outcomes. We further compare simulated interactions with real food-recommendation A/B tests to assess correlation with historical A/B tests and real engagement metrics. A small number of analyses use an LLM judge. For the subjective human-likeness score, we additionally compare model rankings with human pairwise preferences in Appendix~\ref{sec:win_rate}. For consistency-based analyses, the LLM judge is given explicit evidence, such as the persona, context, interaction history, rationale, and action, and is asked to assess whether they are aligned. Most results rely on observed user behavior such as ratings, actions, timestamps, production metrics, and engagement metrics, as summarized in Table~\ref{tab:evaluation_grounding}. Future work should further validate these judge-based analyses with larger-scale human annotations and report human--LLM agreement across different evaluator models.

\noindent Finally, our evaluation demonstrates strong improvements on domains with moderate temporal depth and session-like interactions. Deploying ContextSim in longer-horizon settings, such as continuous media feeds or mobile app usage, may require additional components, including persistent preference evolution or hierarchical goal management. This direction is especially important if the simulator is used not only to compare fixed recommenders, but also to optimize adaptive policies that influence users over time.

\subsection{Running Time and Cost}
We compare the running time of \textsc{ContextSim} and \textsc{SimUSER} for 1{,}000 user interactions. As reported in \textsc{SimUSER}, it performs API calls to GPT-4o-mini and requires 10.1h for 1{,}000 interactions without parallelization, corresponding to approximately \$16--\$21 in API cost under current pricing. In contrast, \textsc{ContextSim} primarily performs inference with a locally served Qwen3-8B policy (vLLM). Using 4-GPU, this yields an estimated runtime of $\approx$1.3h for 1{,}000 interactions, corresponding to roughly \$13--\$16 in GPU time. Overall, \textsc{ContextSim} is faster than GPT-based simulators and offers competitive or lower cost while avoiding per-call API overhead. In addition, our method is privacy-preserving as all interactions are simulated locally without transmitting user data to external APIs.

\section{Datasets}
\noindent\textbf{MovieLens-1M.} MovieLens-1M is a widely adopted benchmark for recommender-systems research. It contains approximately one million explicit ratings on a 1--5 star scale, collected from 6{,}040 users across 3{,}706 movies. The dataset also provides auxiliary information, including movie titles and genre annotations, as well as basic user demographics such as age, gender, and occupation.

\noindent\textbf{Steam.} The Steam dataset comprises user--game interaction records from the Steam platform. It includes user and game identifiers together with English-language user reviews, and provides game-level metadata such as titles.

\noindent\textbf{AmazonBook.} AmazonBook is a subset of the Amazon product reviews corpus limited to the Books category. It consists of user--item interactions in the form of ratings and textual reviews, accompanied by item metadata including book titles and category labels.

\noindent\textbf{OPeRA.} OPeRA is a dataset developed to evaluate large language models for simulating human online shopping behavior. It contains real-world shopping sessions that integrate survey-based persona information, observations of webpage content, fine-grained user actions (e.g., clicks and navigation), and self-reported rationales explaining user decisions.

\section{Simulation Environment}
Our simulator mirrors real-world recommendation platforms like Netflix and Steam, functioning in a page-by-page manner. Users are initially presented with a list of item recommendations on each page: (i) recommendations for MovieLens, Steam, and AmazonBook, or (ii) a web-shopping page for OPeRA. The recommendation algorithm is structured as a standalone module, allowing including any algorithm. This design features pre-implemented collaborative filtering-based strategies, including random, most popular, Matrix Factorization, LightGCN, and MultVAE. 

\noindent Namely, in \textbf{recommendation domains}, the environment displays a \emph{page} of $M$ recommended items as a single text state $s_t$. For each item, the state includes its title and an item description. The short description is either taken from available domain metadata (when present) or retrieved from the title. If the agent clicks on an item, the simulator reveals a more detailed description for that item in the next state. \noindent We format each page as:
\begin{tcolorbox}[colframe=blue!40!black, colback=white, title=Page Format (Recommendation Domains),breakable]
PAGE \{page\_number\}\\
CONTEXT: \{user\_context\} \\
$<$$-$ \{item\_title\} $-$$>$  $<$$-$ History ratings: \{item\_rating\} $-$$>$ $<$$-$ Summary: \{item\_description\} $-$$>$\\
$<$$-$ \{item\_title\} $-$$>$  $<$$-$ History ratings: \{item\_rating\} $-$$>$ $<$$-$ Summary: \{item\_description\} $-$$>$\\
\ldots
\end{tcolorbox}

\noindent Here, \{user\_context\} is the contextual scenario used by ContextSim (temporal, spatial, situational, goal, and constraint factors). \{item\_rating\} is the agent’s own historical rating when available; otherwise, it defaults to a dataset-derived statistic (e.g., global mean rating).

\noindent The environment supports the following actions: \texttt{[NEXT\_PAGE]}: advance to page $(\texttt{page\_number}+1)$. \texttt{[PREVIOUS\_PAGE]}: go back to page $(\texttt{page\_number}-1)$ when $\texttt{page\_number}>1$.\texttt{[CLICK\_ITEM:<item\_id>]}: show the detailed description for the selected item in the next state, \texttt{[SEARCH]}: given query search for specific items, \texttt{[RATE]}, and \texttt{[EXIT]}: terminate the session.

\noindent In \textbf{web-shopping domains} like OPeRA, each state includes (i) page context, (ii) a product list with attributes that appear in the observation, and (iii) a list of interactive elements identified by semantic IDs:
\begin{tcolorbox}[colframe=blue!40!black, colback=white, title=Page Format (Web-Shopping Domains),breakable]
PAGE \{page\_number\}\\
USER CONTEXT: \{user\_context\}\\
CONTEXT: \{page\_context\}\\
PRODUCTS:\\
$<$$-$ \{product\_title\} $-$$>$ $<$$-$ Price: \{price\} $-$$>$ $<$$-$ Details: \{short\_description\} $-$$>$\\
$<$$-$ \{product\_title\} $-$$>$ $<$$-$ Price: \{price\} $-$$>$ $<$$-$ Details: \{short\_description\} $-$$>$\\
\ldots\\
INTERACTIVE ELEMENTS (semantic IDs):\\
\{semantic\_id\_1\}, \{semantic\_id\_2\}, \ldots, \{semantic\_id\_L\}
\end{tcolorbox}
\noindent where the page context features the web-shopping information exposed by OPeRA (e.g., the current page type such as search, cart contents, cart price, and other page-specific cues). Actions follow the same action space as defined in the OPeRA dataset \cite{wang2025opera}, extended with the navigation actions described above.

\section{Prompts}
\begin{figure*}[tbp]
    \begin{center}
        \includegraphics[width=0.8\linewidth]{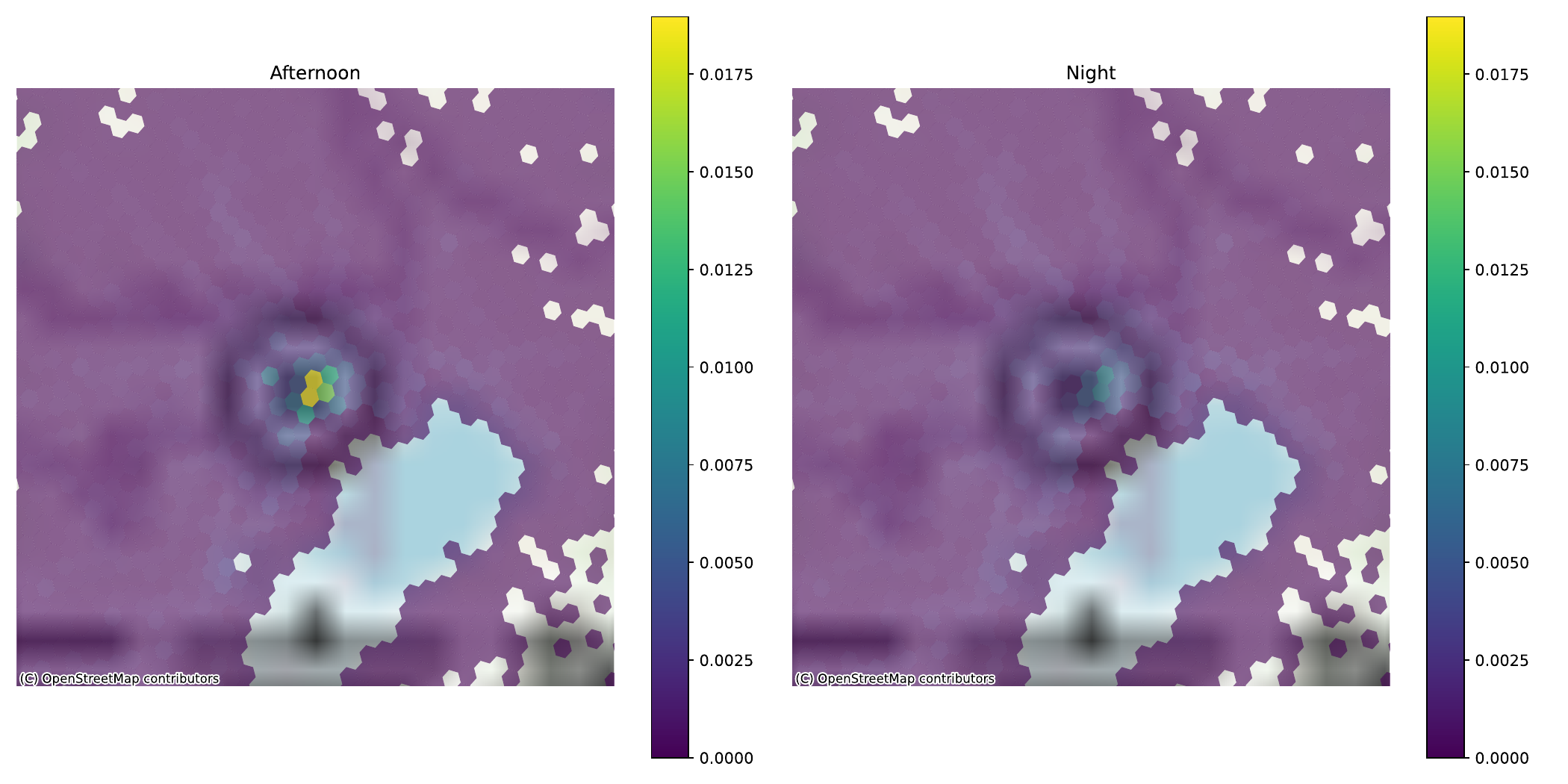}
    \caption{Estimated probability of user interactions with the recommender system across H3 tiles in the Tokyo metropolitan area during the afternoon (left) and night (right). Ocean areas are represented in light blue.}
    \label{fig:location}
\end{center}
\end{figure*}

\subsection{Post-Interview Prompt}
After completing an interaction session, each agent is queried using a post-interview prompt designed to assess its overall satisfaction with the recommender system. The exact prompt used for this evaluation is presented below:
\begin{tcolorbox}[colframe=blue!40!black, colback=white, title=Post-Interview Prompt, breakable]
How satisfied are you with the recommender system you recently interacted with?\\
\textbf{\#\#\# Instructions:}\\
1. Rating: Provide a rating from 1 to 10.\\
2. Explanation: Explain the reason for your rating.\\

\textbf{\#\#\# Response Format:}\\ 
- RATING: [integer between 1 and 10]\\
- REASON: [detailed explanation]\\
\end{tcolorbox}

\subsection{Believability of Synthetic User Prompt}
To evaluate the believability of synthetic users, as described in Section~\ref{sec:beliaviability}, we adapt the post-interview setup by introducing additional instructions focused on prior interactions with recommended items. The modified prompt is provided verbatim below:
\begin{tcolorbox}[colframe=blue!40!black, colback=white, title=Believably of Synthetic User Prompt, breakable]
\textbf{\#\#\# Instructions}

1. Review each \textcolor{customorange}{\{item\_type\}} in the \#\# Recommended List \#\#.\\
2. For each \textcolor{customorange}{\{item\_type\}}, classify if you have already interacted with it (``Interacted'') or if you have not (``Not Interacted'').
\end{tcolorbox}

\subsection{LLM Evaluator Prompt}
To assess whether interaction traces resemble those of real users or are indicative of AI-generated behavior, we employ an external LLM-based evaluator. This evaluator is prompted as follows:
\begin{tcolorbox}[colframe=blue!40!black, colback=white, title=LLM Evaluator Prompt, breakable]
Please evaluate the following interactions of an agent with a recommender system, and determine whether it is generated by a Large Language Model (LLM) AI or a real human:\\
\textcolor{customorange}{\{interaction logs\}}\\

Please rate on a scale of 1 to 5, with 1 being most like an AI and 5 being most like a human. 
\end{tcolorbox}

\section{Additional Experiments}

\subsection{Ablation Studies}
\definecolor{myblue}{RGB}{2,125,181}
\definecolor{myRed}{RGB}{255,123,122}
\begin{table}[tbp]
\centering
\resizebox{1.0\linewidth}{!}{
\begin{tabular}{lcccc}
\toprule
\textbf{Configuration} & \textbf{Accuracy (1:1)} \textcolor{myRed}{$\vardiamondsuit$} & \textbf{RMSE} \textcolor{myRed}{$\vardiamondsuit$} & \textbf{Consistency} \textcolor{myblue}{$\spadesuit$} & \textbf{Temporal Correlation } \textcolor{myRed}{$\vardiamondsuit$} \\
\midrule
ContextSim (full) & 0.824 & 0.451 & 84.1\% & 0.94 \\
\phantom{  } - Life Simulation & 0.798 & 0.489 & 63.8\% & 0.31 \\
\phantom{  } - Thought Synthesis & 0.812 & 0.468 & 46.4\% & 0.88 \\
\bottomrule
\end{tabular}
}
\caption{Component ablation study. We report user preference accuracy, rating RMSE, thought consistency, and temporal correlation, on MovieLens \textcolor{myRed}{$\vardiamondsuit$} and OPeRA \textcolor{myblue}{$\spadesuit$}.}
\label{tab:ablation}
\end{table}
We report ablation results in Table~\ref{tab:ablation}, evaluating preference accuracy, rating error, consistency, and temporal alignment. Removing the life simulation module leads to a sharp drop in temporal correlation, indicating that realistic daily schedules are essential for reproducing time-of-day interaction patterns. In contrast, ablating the thought-synthesis module preserves overall accuracy and RMSE but substantially degrades internal consistency, showing that explicit reasoning is critical for generating coherent, persona-aligned trajectories. Together, these results demonstrate that both components contribute complementary aspects of fidelity: life simulation governs \emph{when} users act, whereas thought synthesis governs \emph{how} they behave. 

\subsection{Interaction Location}

Next, we examine \textit{where} users are likely to engage with recommender systems. We simulate life in the Tokyo area and construct spatial probability maps over H3 tiles, indicating the likelihood that a user will engage with the RS. We report results in Figure \ref{fig:location}. Afternoon interactions (12--18) are sharply concentrated around the central business district, reflecting well-known commuting patterns and the higher density of workplaces. In contrast, nighttime interactions (00--06) remain centered in Tokyo but also spread outward into suburban and residential areas. This shift highlights that user attention is shaped not only by temporal context but also by the mobility structure of daily life, impacting their potential needs and time constraints.

\subsection{LLM Backbone Choice}
\begin{table}[btp]
\centering
\resizebox{1.0\columnwidth}{!}{
\begin{tabular}{lcccc}
\toprule
\textbf{Backbone} & \textbf{MovieLens} & \textbf{AmazonBook} & \textbf{Steam} & \textbf{OPeRA} \\
\midrule
ContextSim (Llama-3.2-3B) \llmlogo{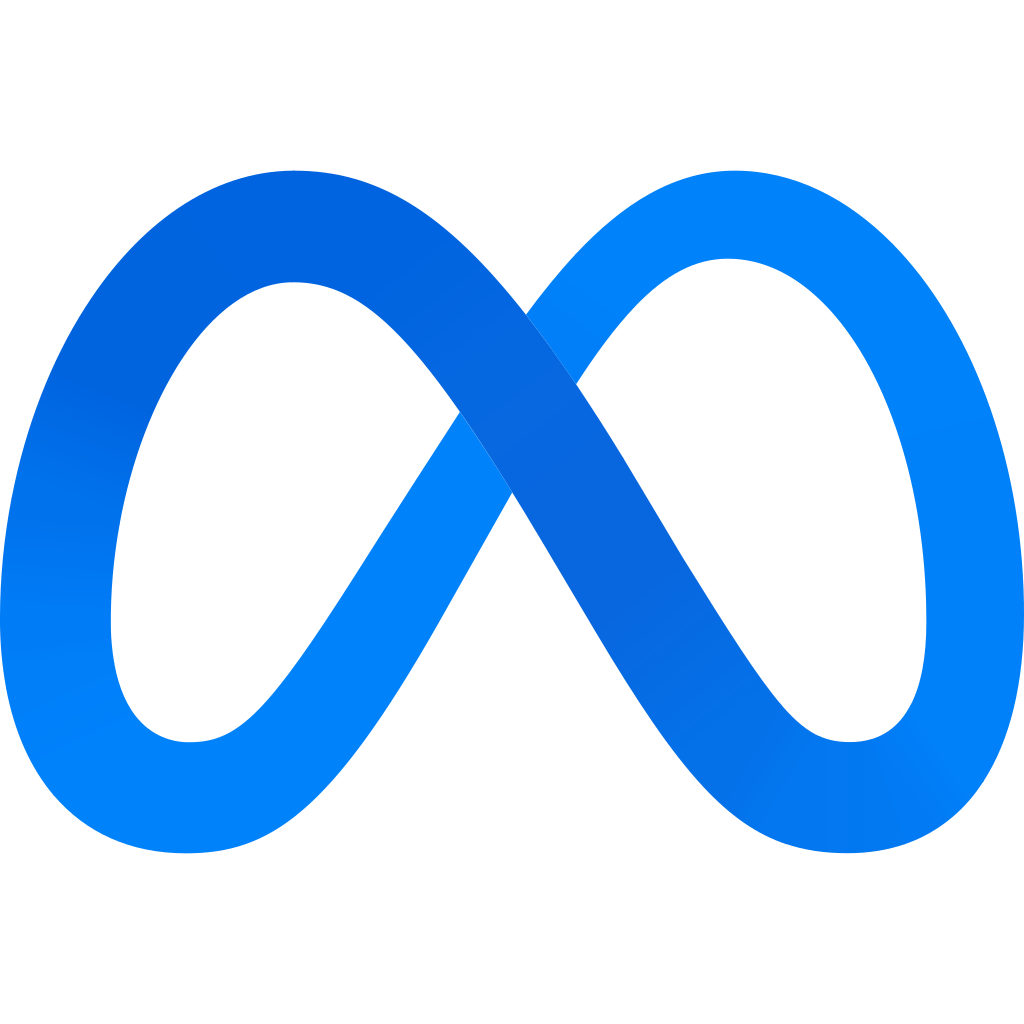}  & 4.15 $\pm$ 0.18 & 4.00 $\pm$ 0.19 & 4.02 $\pm$ 0.21 & 3.96 $\pm$ 0.20 \\
ContextSim (Qwen-2.5-7B) \llmlogo{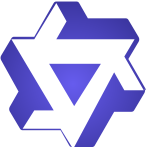}   & 4.35 $\pm$ 0.16 & 4.10 $\pm$ 0.18 & 4.12 $\pm$ 0.20 & 4.03 $\pm$ 0.19 \\
ContextSim (Llama-3.1-8B) \llmlogo{meta-color.png}  & 4.47 $\pm$ 0.15 & 4.20 $\pm$ 0.17 & 4.24 $\pm$ 0.20 & 4.12 $\pm$ 0.18 \\
\rowcolor{blue!10}
ContextSim (Qwen3-8B)  \llmlogo{Qwen_logo.png}     & \textbf{4.60 $\pm$ 0.14} & \textbf{4.28 $\pm$ 0.16} & \textbf{4.31 $\pm$ 0.19} & \textbf{4.21 $\pm$ 0.18} \\
\bottomrule
\end{tabular}}
\caption{Human-likeness scores of \textsc{ContextSim} with different backbone LLMs, evaluated by GPT-4o across four recommendation domains.}
\label{tab:backbone_believability}
\end{table}
To study the impact of the backbone LLM, the base model is swapped while keeping the rest of \textsc{ContextSim} unchanged. As shown in Table~\ref{tab:backbone_believability}, all backbones achieve high human-likeness scores ($\ge4$), indicating that the proposed life-context grounding and thought synthesis are robust to the choice of underlying model. Qwen3-8B yields the strongest results overall, but the performance gaps to Llama-3.1-8B and Qwen-2.5-7B are modest, suggesting that our approach is reasonably robust to the backbone choice.

\subsection{Impact of Contextual Factors}
\begin{table}[tbp]
\centering
\resizebox{1.0\columnwidth}{!}{
\begin{tabular}{lcc}
\toprule
\textbf{Context Factors} & \textbf{Rating error} & \textbf{Temporal correlation} \\
\midrule
None & 0.489 & 0.23 \\
Time only & 0.472 & 0.71 \\
Time + Location & 0.464 & 0.82 \\
Time + Location + Situation & 0.458 & 0.89 \\
Full context & 0.451 & 0.94 \\
\bottomrule
\end{tabular}}
\caption{Impact of individual context factors on MovieLens, with rating error measured by RMSE.}
\label{tab:context_ablation}
\end{table}
We further measure which contextual signals matter most. Table~\ref{tab:context_ablation} highlights how each additional factor improves both rating accuracy and temporal alignment (MovieLens). Using only temporal context already yields a substantial gain in temporal. Adding location and mood progressively closes the gap, with full context achieving the best RMSE and temporal correlation. These results indicate that both when users interact (time), where they are (location), and why they act jointly shape realistic interaction patterns. As expected, rating accuracy is primarily driven by user preferences, which explains the comparatively smaller improvements from contextual factors alone.

\subsection{Action Alignment}
\begin{table*}[tbp]
\small
\centering
\begin{tabular}{lcccc}
\toprule
\textbf{Model} &
\makecell{Action Gen.\\(Accuracy)} &
\makecell{Action Type\\(Macro F1)} & 
\makecell{Click Type\\(Weighted F1)} &
\makecell{Session Outcome\\(Weighted F1)} \\
\midrule
GPT-4.1 \llmlogo{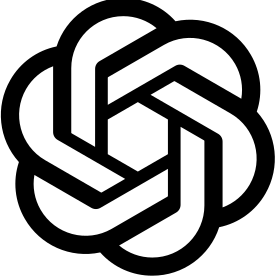} & 21.51  & 48.78 & 44.47 & 47.54 \\
\quad w/o persona & 22.06  & 45.55 & 43.45 & 58.47 \\
\qquad w/o rationale & 21.28 & 34.93 & 42.63 & 51.17 \\
\midrule
Claude-3.7 \llmlogo{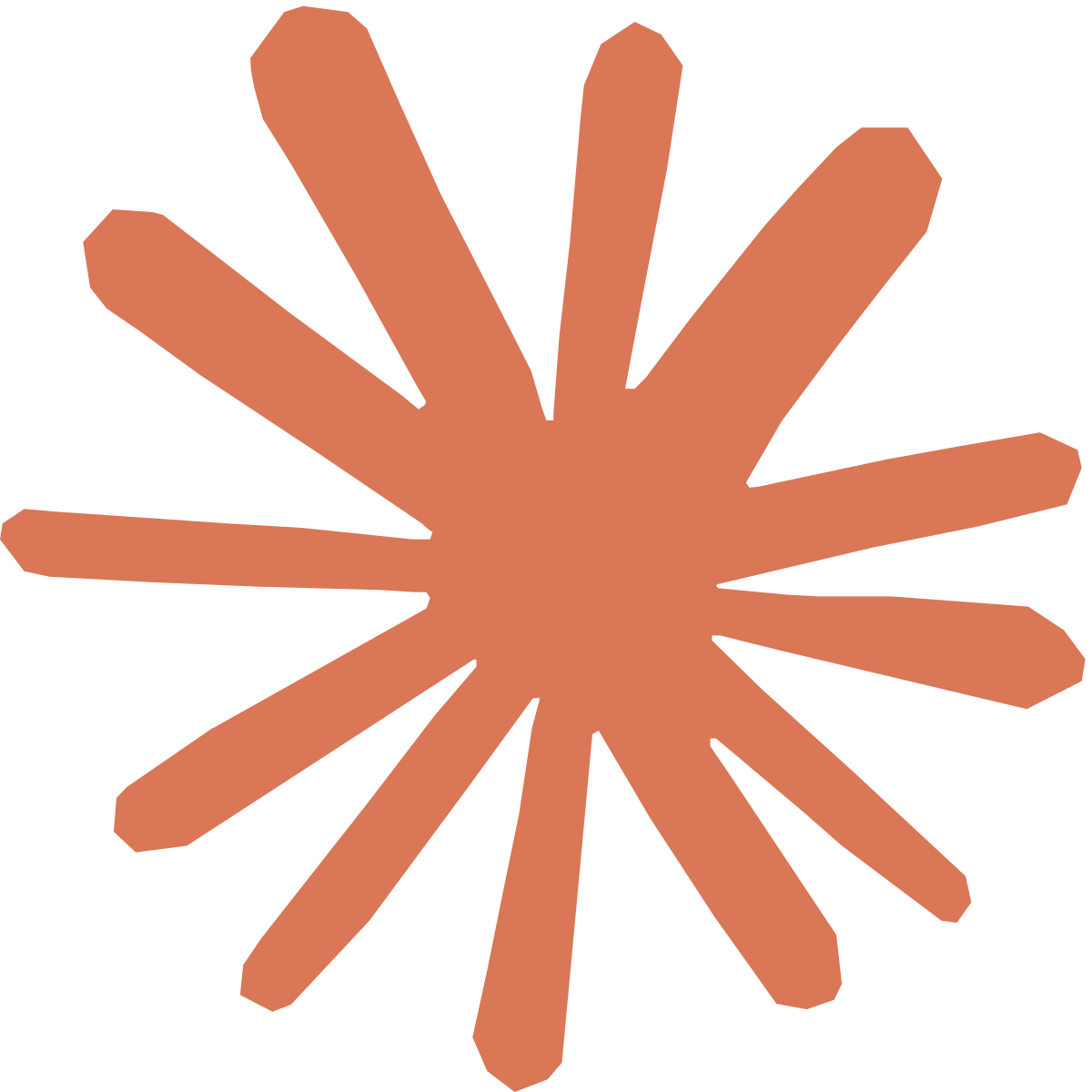} & 10.75  & 31.58 & 27.27  & 43.52 \\
\quad w/o persona & 10.75  & 25.33 & 22.76 & 43.10 \\
\qquad w/o rationale & 10.08  & 26.06 & 20.29  & 43.10 \\
\midrule
Llama-3.3  \llmlogo{meta-color.png} & 8.31  & 24.29 & 19.99  & 36.64 \\
\quad w/o persona & 8.31  & 23.69 & 18.59  & 33.21 \\
\qquad w/o rationale & 8.76  & 23.60 & 19.22 & 34.19 \\
\midrule
RecAgent  \llmlogo{OpenAI_Logo.png} & 22.71 & 49.18 & 45.25 & 54.12 \\
Agent4Rec  \llmlogo{OpenAI_Logo.png} & 23.09 & 50.05 & 46.37 & 56.70 \\
SimUSER  \llmlogo{OpenAI_Logo.png} & 24.21 & 52.44 & 48.68 & 59.63 \\
\rowcolor{blue!10}
ContextSim \llmlogo{Qwen_logo.png} & \textbf{35.37} & \textbf{64.22} & \textbf{61.03} & \textbf{73.74} \\
\bottomrule
\end{tabular}
\caption{
Evaluation of next-action prediction. We report four metrics: action generation accuracy, action type macro F1, click type weighted F1, and session outcome weighted F1. ``Claude-3.7'' denotes Claude-3.7-Sonnet; ``Llama-3.3'' denotes Llama-3.3-70B-Instruct. All metrics are percentages (\%).
}
\label{tab:action_predict}
\end{table*}
Using the OPeRA dataset, action alignment is measured with an exact-match criterion: a prediction is counted as correct only if all action parameters match the ground-truth specification. For \texttt{click} actions, this requires matching the clicked target (e.g., the correct product or button). For \texttt{input} actions, the model must identify both the appropriate input field and the exact text entered by the user. We also assess how well each approach classifies action types. We report F1 scores for the high-level action categories \texttt{click}, \texttt{input}, and \texttt{terminate}. To probe fine-grained behavior, we further compute weighted F1 over \texttt{click} subtypes, capturing whether the model can distinguish between different click intents (e.g., \texttt{review}, \texttt{product\_link}, \texttt{purchase}). Given that online shopping is inherently goal-driven, we evaluate the prediction of session outcomes. Each session terminates either with a \texttt{purchase-related click} or a \texttt{terminate} action. Performance on these terminal actions is assessed using accuracy and weighted F1, reflecting how well the model captures users’ final decisions and long-term goals over the course of a session. As presented in Table~\ref{tab:action_predict}, \textsc{ContextSim} substantially outperforms the baselines, with especially large gains in action generation accuracy and session-outcome prediction.

\subsection{Context Consistency}
\begin{table}[t]
\centering
\resizebox{0.95\columnwidth}{!}{
\begin{tabular}{lccc}
\toprule
\textbf{Context factor} & \textbf{Aligned} & \textbf{Partially aligned} & \textbf{Contradictory} \\
\midrule
Temporal $c_t$    & 0.69 & 0.23 & 0.08\\
Spatial $c_l$     & 0.64 & 0.27 & 0.09 \\
Situational $c_s$ & 0.55 & 0.32 & 0.13 \\
Goal $c_g$        & 0.72 & 0.21 & 0.07 \\
Constraint $c_b$  & 0.58 & 0.30 & 0.12 \\
\bottomrule
\end{tabular}}
\caption{LLM-based consistency between simulated contexts and OPeRA survey answers. Results are averaged over 5{,}000 interaction samples.}
\label{tab:context_consistency}
\end{table}
We next examine whether the contextual factors are consistent with real-world contexts provided in OPeRA~\cite{wang2025opera}. We run the daily-life module for 30 days, and collect the aggregated $c$. For evaluation, we randomly sample 5{,}000 interaction points across all agents. For each sampled context and each dimension (temporal, spatial, situational, goal, constraint), we provide an LLM judge with: (i) the user’s persona, (ii) the textual description of the simulated context, and (iii) the ground truth from OPeRA. The judge is asked to label, independently for each dimension, whether the simulated context is \emph{aligned}, \emph{partially aligned}, or \emph{contradictory} to the survey answers. As illustrated in Table~\ref{tab:context_consistency}, contextual factors are largely coherent with the ground-truth. Temporal and goal-related factors exhibit the highest alignment, while situational and constraint factors show slightly more variability but remain largely consistent.

\subsection{Rating Distribution}
\begin{figure}[tbp]
    \centering
    \includegraphics[width=0.90\linewidth]{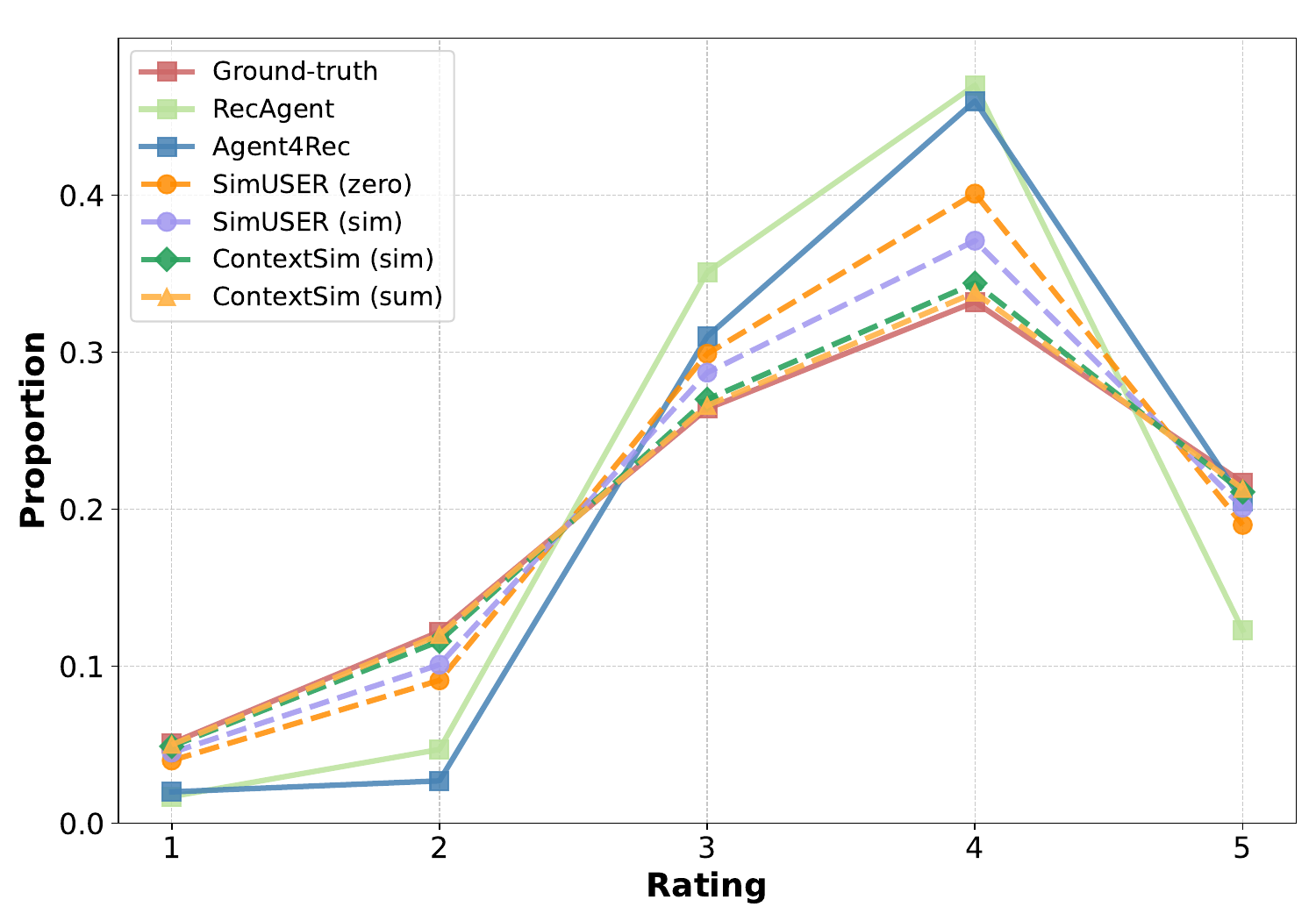}
    \caption{Comparison of rating distributions between ground-truth and human proxies.}
    \label{fig:rating_alignement}
\end{figure}

Beyond individual rating alignment, human proxies must replicate real-world behavior at the macro level. This implies ensuring that the distribution of ratings generated by the agents matches the distributions observed in the original dataset. Figure~\ref{fig:rating_alignement} presents the rating distribution from the MovieLens dataset and the ratings generated by different simulators. These results reveal a high degree of alignment between the simulated and actual rating distributions, with a predominant number of ratings at 4 and a small number of low ratings (1--2). While Agent4Rec and SimUSER assign fewer low ratings than real users, \textsc{ContextSim(sim)} and \textsc{ContextSim(sum)} more closely match the true distribution, indicating improved alignment at the population level.

\subsection{Time-Aware Interaction Patterns}
\begin{table}[tbp]
\centering
\resizebox{1.0\linewidth}{!}{
\begin{tabular}{lcccc}
\toprule
\textbf{Time Period} & \textbf{Real} & \textbf{Agent4Rec} & \textbf{SimUSER} & \textbf{ContextSim(sim)} \\
\midrule
Morning (6-12) & 0.11 & 0.25 & 0.25 & 0.16 \\
Afternoon (12-18) & 0.21 & 0.25 & 0.25 & 0.24 \\
Evening (18-24) & 0.35 & 0.25 & 0.25 & 0.37 \\
Night (0-6) & 0.28 & 0.25 & 0.25 & 0.23 \\
\bottomrule
\end{tabular}
}
\caption{Temporal click-through rate patterns. }
\label{tab:temporal_patterns}
\end{table}

In this study, we postulate capturing contextual patterns present in real user behavior but absent from context-free simulators. Using MovieLens interactions with timestamps, we group clicks into four time-of-day bands and compare temporal click-through rates. Table \ref{tab:temporal_patterns} demonstrates that \textsc{ContextSim(sim)} accurately captures the evening peak in engagement and reduced night activity, while baseline methods use unrealistic uniform patterns. The correlation between ContextSim and real temporal patterns is higher compared to SimUSER.

\subsection{Brand Loyalty}
\label{sec:brand_loyalty}

\begin{figure}[t]
    \centering
    \includegraphics[width=0.9\linewidth]{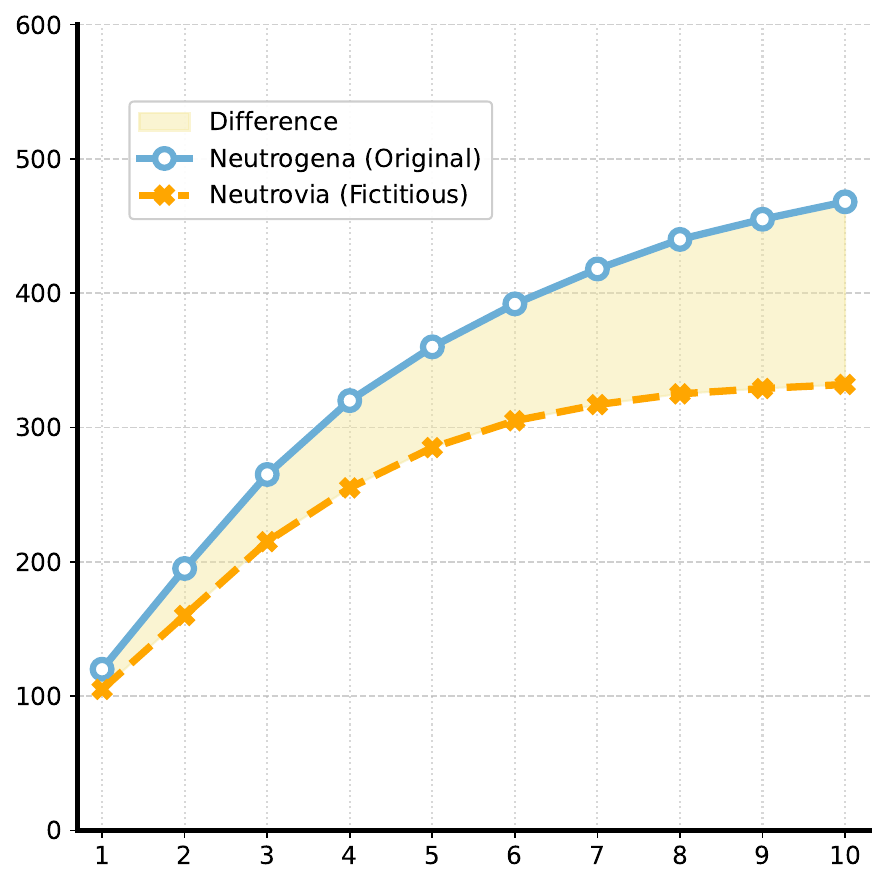}
    \caption{The phenomenon observation of Brand Loyalty.}
    \label{fig:brand_loyalty_curve}
\end{figure}

Next, we analyze the proportion of interactions each item received at the final stage of the simulation relative to the total number of interactions, as reported in Figure~\ref{fig:brand_loyalty_curve}. Because our evaluation includes web-shopping trajectories (OPeRA), where items explicitly expose brand names (e.g., titles and product-page content), brand cues can directly influence agent decisions. We noticed that brand-identified items were significantly more popular than their counterparts. To further examine this effect, we replace the brand name \textit{``Neutrogena''} with a fictitious alternative, \textit{``Neutrovia''}, while keeping all other item attributes unchanged, including recommendation probability and non-brand content. Here, a step corresponds to one interaction round in the simulator. Figure~\ref{fig:brand_loyalty_curve} suggests that removing the recognizable brand cue substantially reduces engagement: after 10 steps, the original-brand item accumulates 468 likes, compared to 332 for the fictitious-brand variant.

\subsection{Matthew Effect}
\label{sec:matthew_effect}

\begin{figure}[t]
    \centering
    \includegraphics[width=0.9\linewidth]{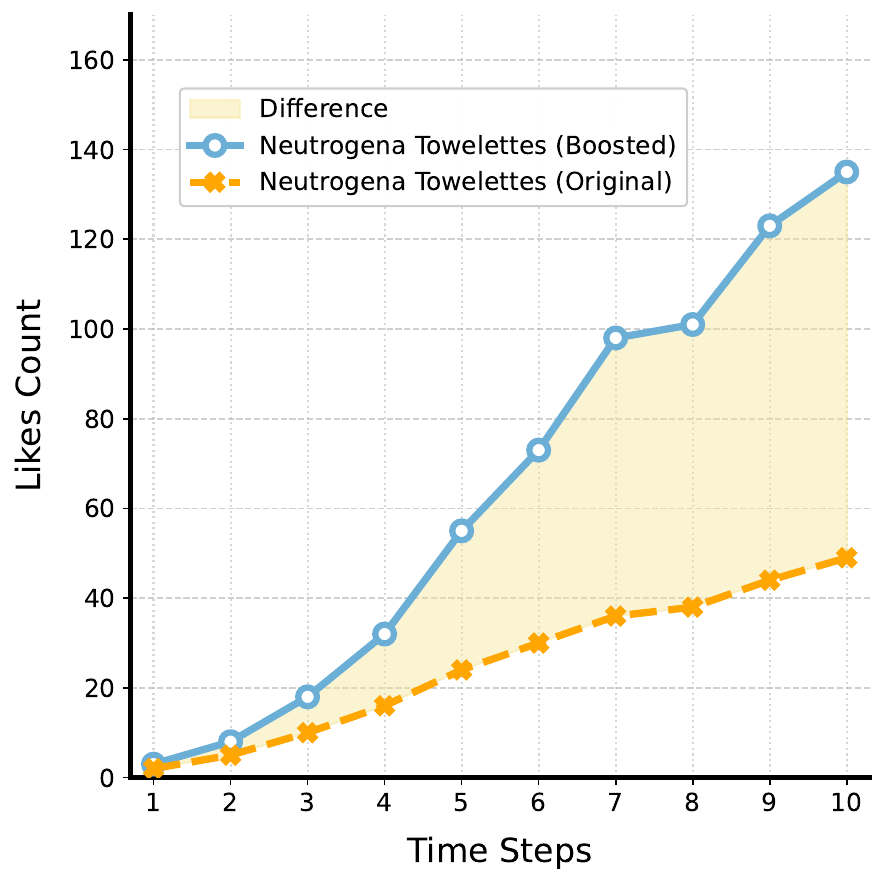}
    \caption{The phenomenon observation of Matthew effect.}
    \label{fig:mathew}
\end{figure}
We quantify the \emph{Matthew Effect} \cite{wang2018quantitative} in our simulator by introducing a small early exposure advantage to a single target product and measuring whether this initial advantage compounds over repeated interaction rounds. We select \texttt{Neutrogena Make-Up Remover Cleansing Towelettes}, and evaluate two settings: (i) Original, where the product is ranked according to the recommender as usual, and (ii) Boosted, where the same product receives a slight exposure advantage only during the first two interaction rounds, after which the recommendation policy is identical to the Original condition. We track the cumulative number of positive interactions (``likes'') received by the target product over steps in both conditions. As shown in Figure~\ref{fig:mathew}, a small initial exposure boost results in a persistent and widening gap in cumulative likes, even after the boost is removed, consistent with the Matthew effect in simulated interactions.

\subsection{Context Effects on Preferences}
\label{sec:context_effects}

\begin{figure}[t]
    \centering
    \includegraphics[width=0.95\linewidth]{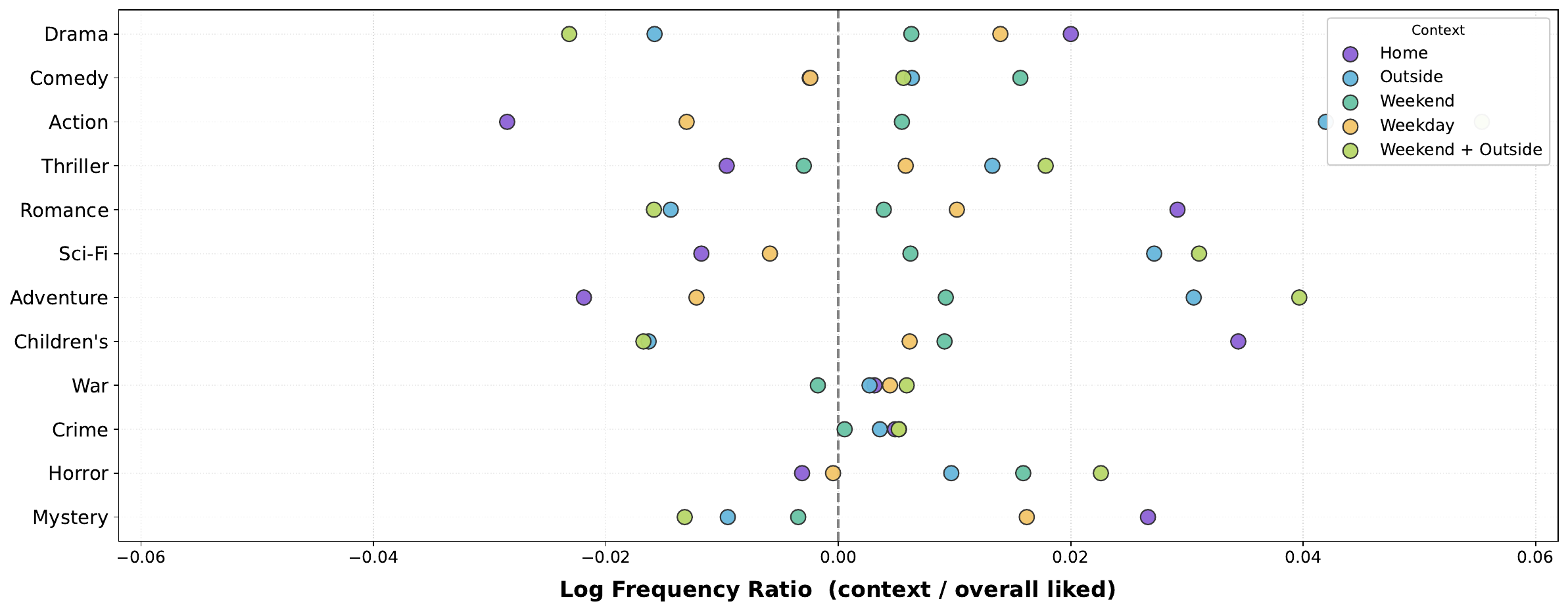}
    \caption{Context effects on agent likes in simulation. Each point shows the log frequency ratio of a genre within a context relative to the overall liked distribution across all contexts.}
    \label{fig:context_effects}
\end{figure}

A core motivation of our framework is that user decisions are shaped by \emph{context} (e.g., being at home vs outside, weekday vs weekend). To test whether our agents respond to context in a behaviorally meaningful way, we analyze how the distribution of \emph{liked} items changes across contexts produced by our life simulation module. We run the simulator on MovieLens items and log each interaction together with the agent's context label (location: home/outside; day type: weekday/weekend) assigned by our daily-life module. We define an interaction as positive (``liked'') when the agent issues a positive feedback event (e.g., rating $\geq 4$). For each context $c$ and genre $g$, we report the log frequency ratio: $\log \frac{p_{\text{sim}}(g \mid c)}{p_{\text{sim}}(g \mid \text{all})}$, where $p_{\text{sim}}(g \mid c)$ is the empirical genre frequency among liked interactions in context $c$. Figure~\ref{fig:context_effects} visualizes these shifts for the most frequent genres. Positive values indicate that a genre is over-represented among liked items in a given context compared to the global average, reflecting a positive contextual bias, while negative values indicate under-representation. For example, \emph{Drama} is more likely to be liked at home, whereas some genres (e.g., \emph{Crime}) show near-zero shifts, suggesting limited sensitivity to contextual factors.

\subsection{Human Pairwise Preference}
\label{sec:win_rate}

\begin{figure}[t]
    \centering
    \includegraphics[width=1.0\linewidth]{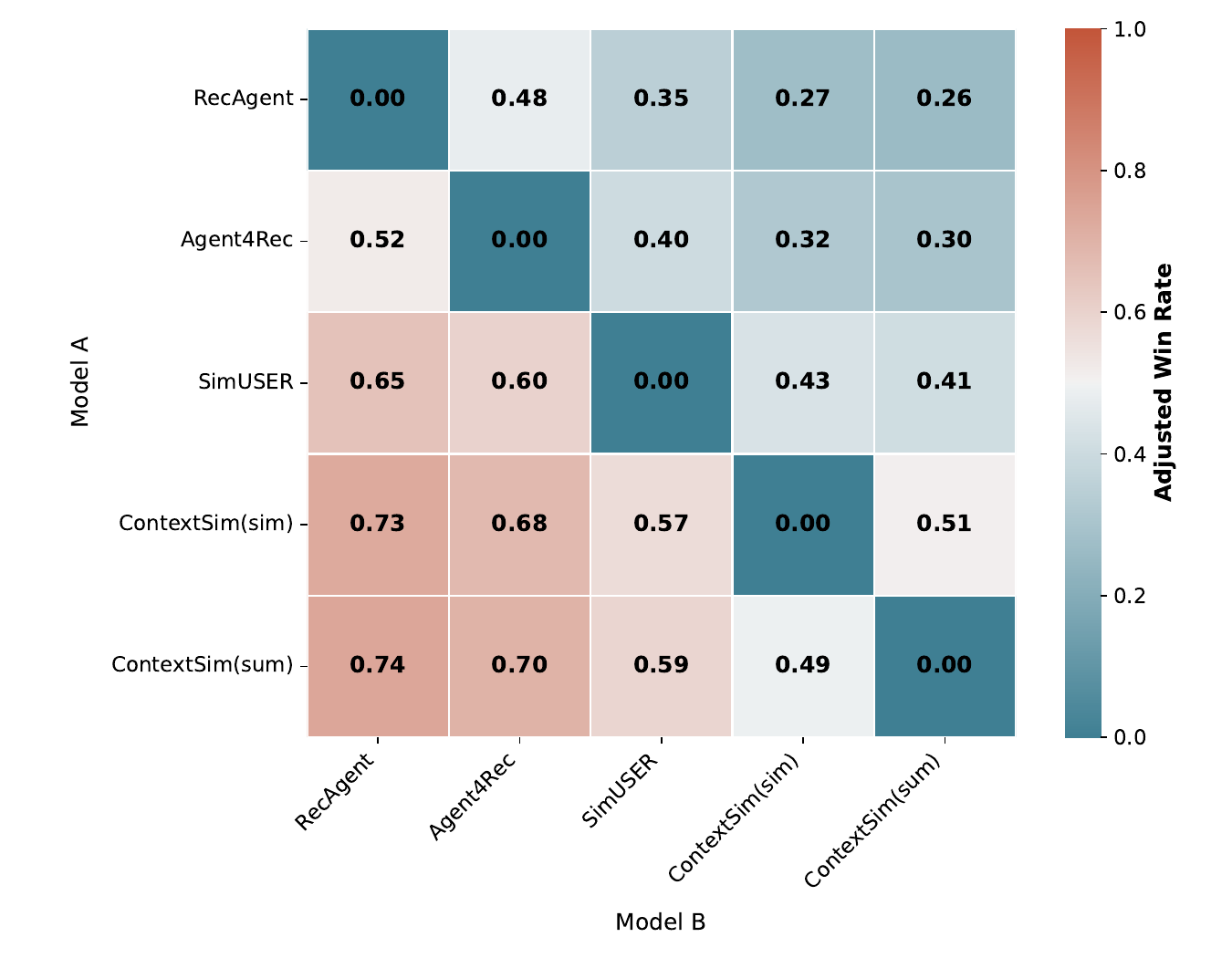}
    \caption{
    Win matrix heatmap based on pairwise preference evaluation. Each entry \( w_{ij} \) denotes the adjusted win probability that method \( i \) is preferred over method \( j \).}
    \label{fig:human_win_matrix}
\end{figure}
To assess the quality of simulator-generated trajectories, five evaluators were given 200 samples, each with two anonymous trajectories generated by two different methods for the same underlying input. The evaluator was tasked with selecting the preferred trajectory between the two provided (ties allowed). We ranked methods using a win matrix (Figure~\ref{fig:human_win_matrix}) and Bradley--Terry (BT) model coefficients. The win matrix records matchup outcomes, where element \( w_{ij} \) indicates the probability that method \( i \) defeats method \( j \), with ties counted as half wins. Overall, the resulting rankings are consistent with those obtained from the LLM-based judges (see section \ref{sec:believability}).

\subsection{Generalization to Unfamiliar Items}
\begin{figure*}[tbp]
    \centering
    \includegraphics[width=0.45\linewidth]{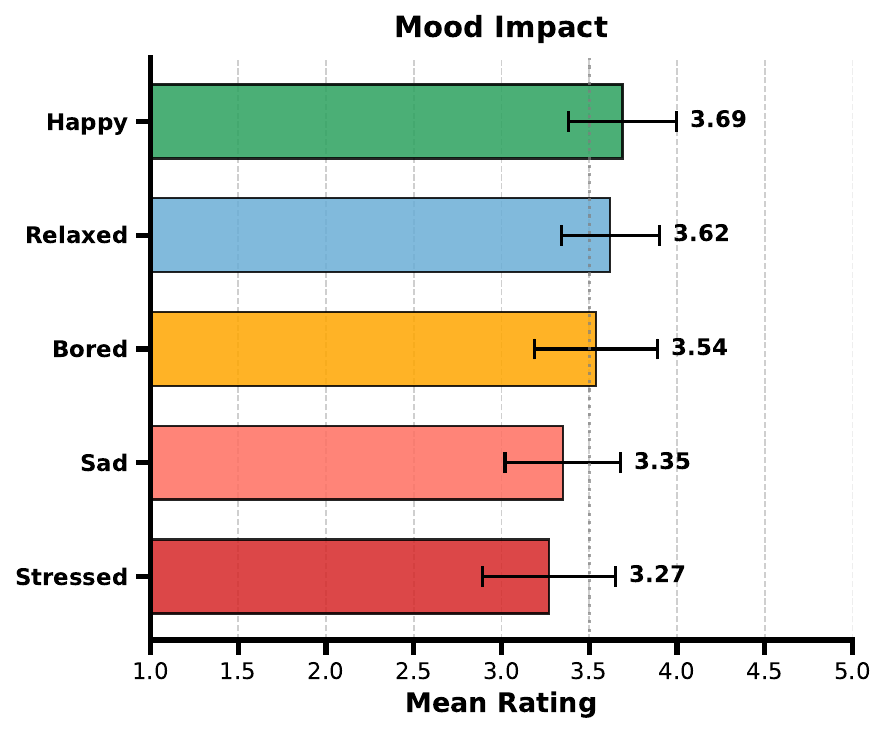}
    \includegraphics[width=0.45\linewidth]{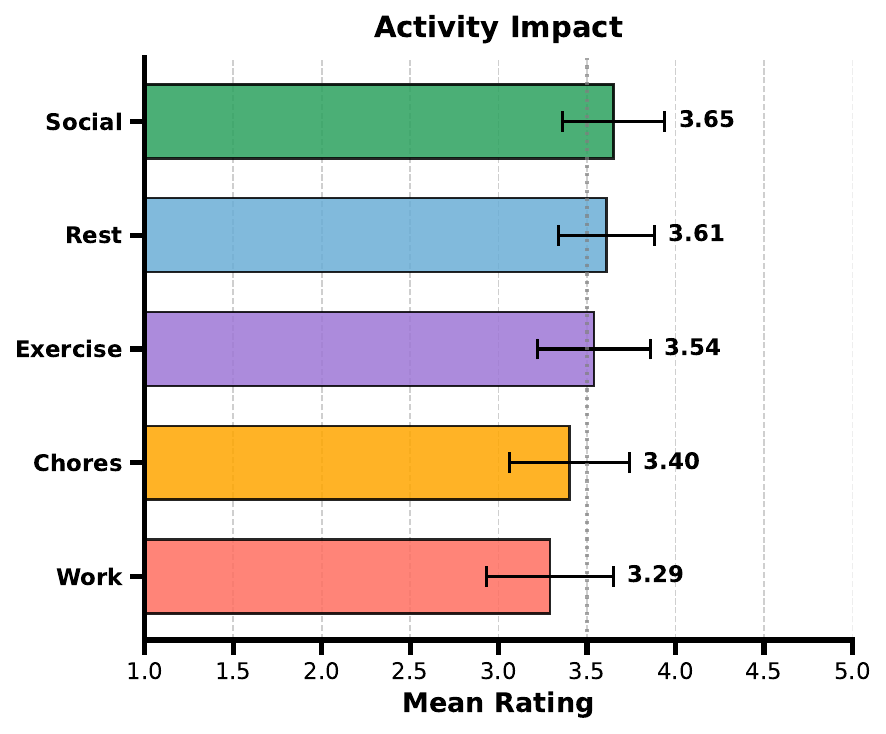}
    \caption{Impact of situational context on predicted ratings on MovieLens.}
    \label{fig:mood_activity_separate}
\end{figure*}
\begin{figure}[tbp]
    \centering
    \includegraphics[width=1.0\linewidth]{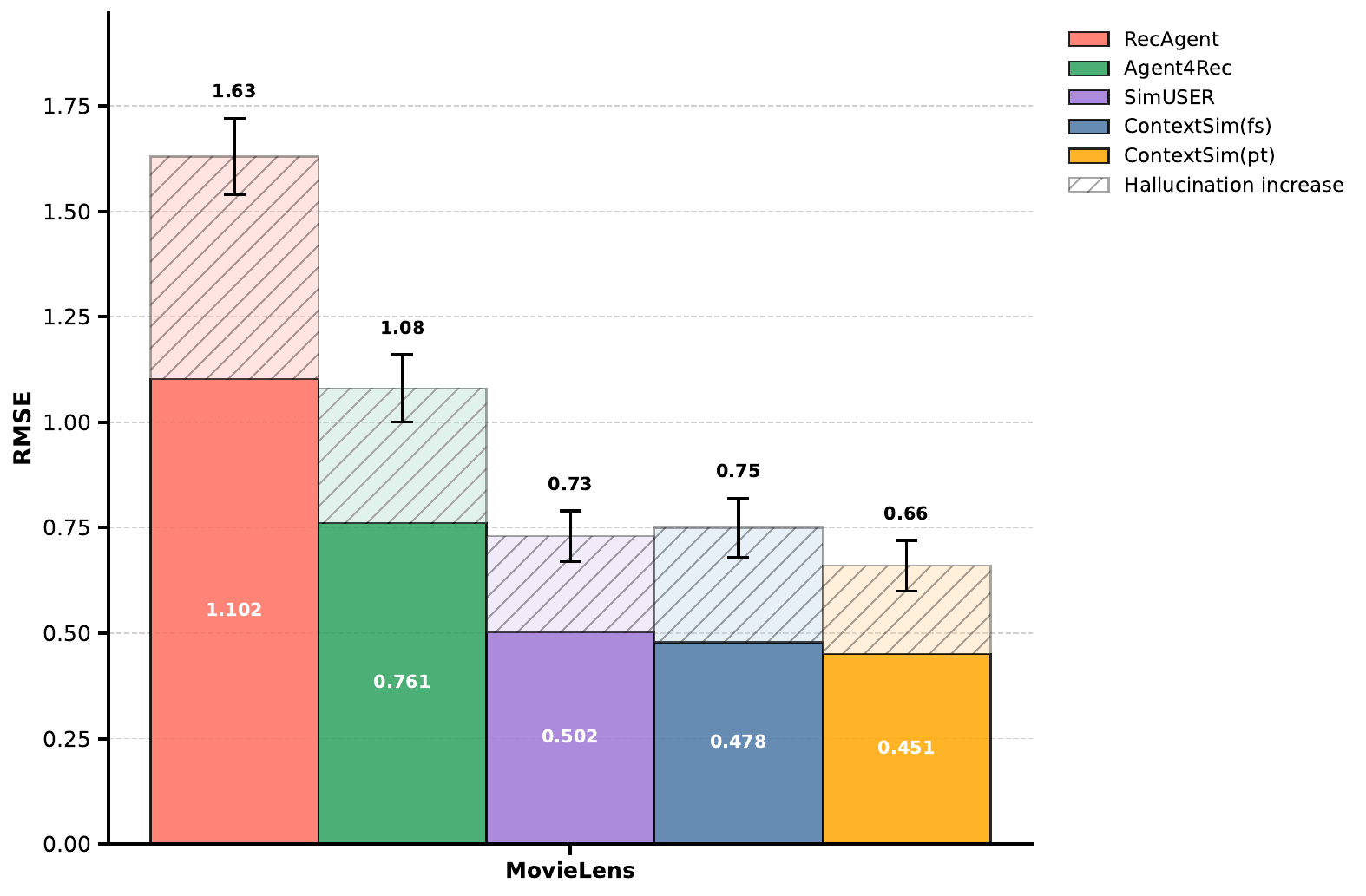}
    \caption{Comparison of RMSE values for the standard rating task (dark bars) and the hallucination subset (dark+light stacked bars) on MovieLens.}
    \label{fig:hallucination_rating}
\end{figure}

In this experiment, we target items that are likely unfamiliar to the backbone LLM, in order to evaluate whether thought synthesis improves generalization beyond the training distribution. Following the rating evaluation in Table~\ref{fig:rating_prediction}, including a few-shot prompting version of our framework, \textsc{ContextSim(fs)}. Concretely, we query the backbone LLM to classify each movie into one of the dataset genres; items whose predicted genre does not match the ground-truth label are treated as unfamiliar and included in the subset, while correctly classified items are excluded. Figure~\ref{fig:hallucination_rating} shows that RMSE increases for all methods under hallucination, as expected. However, \textsc{ContextSim} is the most robust, exhibiting the smallest degradation relative to its original performance. In contrast, the few-shot variant, \textsc{ContextSim(fs)}, degrades more sharply and becomes comparable to \textsc{SimUSER} on unfamiliar items, consistent with the brittleness of few-shot prompting --- when the model cannot anchor decisions in stable learned reasoning. Overall, these results highlight that pretraining improves generalization by teaching the policy to discover preference-relevant attributes and how preferences align with its persona.

\subsection{Impact of Situational Context}

We finally investigate how situational context, specifically mood and recent activities, influences user engagement with recommendations. Using MovieLens, we report the average rating conditioned on each contextual state. As reported in Figure~\ref{fig:mood_activity_separate}, both factors significantly affect predicted ratings. Positive moods (happy, relaxed) increase ratings compared to negative states (stressed, sad), consistent with evaluation documented in psychology literature~\cite{mayer1992mood}. Bored users exhibit moderately high ratings, likely reflecting heightened receptivity to novel entertainment. Similarly, leisure-oriented activities (social gatherings, rest) yield higher ratings than cognitively demanding tasks (work, chores), suggesting users are more receptive to recommendations after relaxation. These findings demonstrate that ContextSim captures nuanced psychological effects that static preference models ignore. By grounding agents in a realistic situational context, our framework enables more accurate simulation of context-dependent user behavior, which is critical for evaluating RS performance across diverse real-world scenarios.

\subsection{Prompt Robustness}
\label{sec:prompt_robustness}

\begin{table}[tbp]
\centering
\small
\resizebox{1.0\linewidth}{!}{
\begin{tabular}{lccc}
\toprule
\textbf{Prompt setting} 
& \textbf{MovieLens RMSE} 
& \textbf{OPeRA Action F1} 
& \textbf{MovieLens Temporal corr.} \\
\midrule
Original prompts 
& 0.451 
& 64.22 
& 0.94 \\
Mean over prompt variants 
& 0.459 
& 63.61 
& 0.91 \\
Worst prompt variant 
& 0.470 
& 62.77 
& 0.89 \\
\bottomrule
\end{tabular}
}
\caption{Prompt robustness analysis. The model is fixed, only prompt wording is varied.}
\label{tab:prompt_robustness}
\end{table}
Since several components of \textsc{ContextSim} are prompt-driven, we examine whether the results depend on a single hand-crafted prompt design. We keep the fine-tuned policy and vary only the wording of prompts used by the persona initialization, life simulation, action selection, and reflection modules. Specifically, we create ten prompt variants that preserve the same task definition but differ in instruction order, phrasing, and the amount of explanation requested from the agent. For each variant, we rerun the simulator on the same subset of agents and recommendation states, using the same random seeds. We then report performance variation across prompt variants. As shown in Table~\ref{tab:prompt_robustness}, performance varies only moderately across prompt variants. This suggests that the gains are not driven by a single fragile prompt template. At the same time, prompt design remains a source of variance, especially for modules that generate context and reflections. We therefore release all prompt templates and report this analysis to make the prompt dependence of the framework explicit.

\end{document}